\documentclass[journal]{IEEEtran}
\ifCLASSINFOpdf
\else
\fi
\usepackage[utf8]{inputenc}
\usepackage{amsmath,amssymb,amsfonts}
\usepackage{algorithmic}
\usepackage{graphicx}
\usepackage{placeins} 
\usepackage{booktabs}
\usepackage{url}
\usepackage{verbatim}
\usepackage{siunitx}
\usepackage{multirow}
\usepackage{array}
\usepackage{pgfplots, pgfplotstable}
\usepackage{tikz}

\usepackage{tabularx}
\newcolumntype{k}{>{\hsize=.6\hsize}X}
\newcolumntype{m}{>{\hsize=.45\hsize}X}
\newcolumntype{s}{>{\hsize=.3\hsize}X}
\usepackage{cleveref}
\usepackage{pdfpages}
\usepackage{enumitem}
\usepackage{soul}
\usepackage[nowarn]{glossaries}
\usepackage{caption}
\usepackage{subcaption}
\usepackage[ruled,vlined]{algorithm2e}
\usepackage[numbers,sort&compress]{natbib}
\newlength\mylen
\newcommand\algoinput[1]{%
  \settowidth\mylen{\KwIn{}}%
  \setlength\hangindent{\mylen}%
  \hspace*{\mylen}#1\\}
\makeatletter
\newcommand{\removelatexerror}{\let\@latex@error\@gobble}
\makeatother
\usetikzlibrary{plotmarks,spy,backgrounds,trees,calc}

\usepackage{authblk}
\DeclareSIUnit{\kmh}{km/h}
\DeclareSIUnit{\dBm}{dBm}
\newlength{\hFig}
\newlength{\wFig}
\newcommand\Tstrut{\rule{0pt}{2.2ex}}         

\definecolor{plot_color1}{RGB}{31, 119, 180}
\definecolor{plot_color2}{RGB}{255, 127, 14}
\definecolor{plot_color3}{RGB}{44, 160, 44}
\definecolor{plot_color4}{RGB}{214, 39, 40}
\definecolor{plot_color5}{RGB}{148, 103, 189}
\definecolor{plot_color6}{RGB}{140, 86, 75}
\definecolor{plot_color7}{RGB}{227, 119, 194}
\definecolor{plot_color8}{RGB}{127, 127, 127}
\definecolor{plot_color9}{RGB}{188, 189, 34}
\definecolor{plot_color10}{RGB}{23, 190, 207}

\usepackage{import}
\DeclareUnicodeCharacter{2212}{-}

\soulregister\cite7
\soulregister\gls7
\soulregister\ref7
\soulregister\pageref7

\newacronym{6g}{6G}{6th generation}
\newacronym{ph}{PH}{Page-Hinkley}
\newacronym{adf}{ADF}{augmented Dickey–Fuller}
\newacronym{agv}{AGV}{automated guided vehicle}
\newacronym{aic}{AIC}{Akaike information criterion}
\newacronym{ai}{AI}{artificial intelligence}
\newacronym{d2d}{D2D}{device-to-device}
\newacronym{hd}{HD}{high-definition}
\newacronym{iid}{i.i.d.}{independent and identically distributed}
\newacronym{pqos}{pQoS}{predictive quality of service}
\newacronym{SHAP}{SHAP}{Shapley additive explanations}
\newacronym{ALE}{ALE}{accumulated local effects}
\newacronym{qos}{QoS}{quality of service}
\newacronym{pra}{PRA}{predictive resource allocation}
\newacronym{tod}{ToD}{teleoperated driving}
\newacronym{v2v}{V2V}{vehicle-to-vehicle}
\newacronym{ap}{AP}{access point}
\newacronym{ue}{UE}{user equipment}
\newacronym{e2e}{E2E}{end-to-end}
\newacronym{mec}{MEC}{mobile edge computing}
\newacronym{ran}{RAN}{radio access network}
\newacronym{rrm}{RRM}{radio resource management}
\newacronym{ml}{ML}{machine learning}
\newacronym{gpu}{GPU}{graphical processing unit}
\newacronym{GPS}{GPS}{global positioning system}
\newacronym{gdpr}{GDPR}{General Data Protection Regulation}
\newacronym{mac}{MAC}{medium access control}
\newacronym{iot}{IoT}{internet of things}
\newacronym{osi}{OSI}{open systems interconnect}
\newacronym{lstm}{LSTM}{long short-term memory}
\newacronym{gb}{GB}{gradient-boosted decision tree}
\newacronym{gru}{GRU}{gated recurrent unit}
\newacronym{dl}{DL}{deep learning}
\newacronym{udp}{UDP}{user datagram protocol}
\newacronym{tcp}{TCP}{transmission control protocol}
\newacronym{pdf}{PDF}{probability density function}
\newacronym{kde}{KDE}{kernel density estimation}
\newacronym{snr}{SNR}{signal-to-noise ratio}
\newacronym{rf}{RF}{random forest}
\newacronym{los}{LOS}{line-of-sight}
\newacronym{nlos}{NLOS}{non line-of-sight}
\newacronym{de}{RE}{resampling error}
\newacronym{kpi}{KPI}{key performance indicator}
\newacronym{ul}{UL}{uplink}
\newacronym{dll}{DL}{downlink}
\newacronym{pca}{PCA}{principal component analysis}
\newacronym{lr}{LR}{linear regression}
\newacronym{rem}{REM}{radio environment map}
\newacronym{sinr}{SINR}{signal-to-interference-plus-noise ratio}
\newacronym{pps}{PPS}{pulse per second}
\newacronym{bsr}{BSR}{buffer status report}
\newacronym{mme}{MME}{mobility management entity}

\newacronym{mape}{MAPE}{mean absolute percentage error}
\newacronym{mae}{MAE}{mean absolute error}
\newacronym{rmse}{RMSE}{root mean square error}
\newacronym{medae}{MedAE}{median absolute error}

\newacronym{mlp}{MLP}{multilayer perceptron}
\newacronym{ann}{ANN}{artificial neural network}
\newacronym{dnn}{DNN}{deep neural network}
\newacronym{relu}{ReLU}{rectified linear unit}
\newacronym{adam}{Adam}{Adam}

\newacronym{gnet}{CE}{consumer equipment}
\newacronym{minipc}{DME}{dedicated measurement equipment}

\newacronym{kl}{\text{KL}}{Kullback-Leibler}

\newacronym{phy}{PHY}{physical layer}
\newacronym{chan}{CHAN}{channel conditions}
\newacronym{bs}{BS}{base station data}
\newacronym{veh}{VEH}{vehicle information}
\newacronym{REM}{REM}{radio environment map}

\newacronym{md}{MD}{modem access}
\newacronym{emd}{EMD}{extended modem access}
\newacronym{mdnet}{MDNET}{modem and network access}
\newacronym{mdrem}{MDREM}{modem access and statistics}
\newacronym{emdnet}{EMDNET}{extended modem and network access}
\newacronym{remnet}{REMNET}{modem, statistics and network access}
\newacronym{devnet}{DEVNET}{vehicle and network access}
\newacronym{dev}{DEV}{full device access}
\newacronym{full}{FULL}{full access}

\newacronym{pdcp}{PDCP}{packet data convergence protocol}
\newacronym{rlc}{RLC}{radio link control}
\newacronym{rrc}{RRC}{radio resource control}

\newacronym{lte}{LTE}{long term evolution}
\newacronym{fdd}{FDD}{frequency-division duplex}
\newacronym{rsrp}{RSRP}{reference signal received power}
\newacronym{rssi}{RSSI}{received signal strength indicator}
\newacronym{rsrq}{RSRQ}{reference signal received quality}
\newacronym{xai}{XAI}{explainable AI}
\glsunset{adam}
    
\makeglossaries

\begin{document}
%

\title{Machine Learning for QoS Prediction in Vehicular Communication: Challenges and Solution Approaches
\thanks{This work was supported by the Federal Ministry of Education and Research (BMBF) of the Federal Republic of Germany as part of the AI4Mobile project (16KIS1170K). The authors alone are responsible for the content of the paper.} 
}

\setlength{\hFig}{6.0cm}
\setlength{\wFig}{8.5cm}
%
%
%

\author[1]{Alexandros Palaios}
\author[2]{Christian L. Vielhaus}
\author[3]{Daniel F. K\"{u}lzer} 
\author[1]{Cara Watermann}
\author[4]{Rodrigo Hernangomez}
\author[5]{Sanket Partani}
\author[1]{Philipp Geuer}
\author[6]{Anton Krause}
\author[5]{Raja Sattiraju}
\author[4]{Martin Kasparick}
\author[6]{Gerhard Fettweis}
\author[2]{Frank H. P. Fitzek}
\author[5]{Hans D. Schotten}
\author[4,7]{Slawomir Sta\'{n}czak}

\affil[1]{Ericsson Research, Germany, \{alex.palaios, cara.watermann, philipp.geuer\}@ericsson.com}
\affil[2]{Deutsche Telekom Chair, Technische Universit\"{a}t Dresden, Germany, \{christian.vielhaus, frank.fitzek\}@tu-dresden.de}
\affil[3]{BMW Group, Germany, daniel.kuelzer@bmwgroup.com}
\affil[4]{Fraunhofer Heinrich Hertz Institute,  Germany, \{rodrigo.hernangomez, martin.kasparick, slawomir.stanczak\}@hhi.fraunhofer.de}
\affil[5]{Technische Universität Kaiserslautern, Germany, \{partani, sattiraju, schotten\}@eit.uni-kl.de}
\affil[6]{Vodafone Chair, Technische Universit\"{a}t Dresden, Germany, \{anton.krause, gerhard.fettweis\}@tu-dresden.de}
\affil[7]{Network Information Theory Group,Technische Universit\"{a}t Berlin, Germany}

\maketitle


\begin{abstract}
As cellular networks evolve towards the \acrlong{6g}, \acrlong{ml} is seen as a key enabling technology to improve the capabilities of the network. Machine learning provides a methodology for predictive systems, which, in turn, can make networks become proactive. This proactive behavior of the network can be leveraged to sustain, for example, a specific \acrlong{qos} requirement. With \acrlong{pqos}, a wide variety of new use cases, both safety- and entertainment-related, are emerging, especially in the automotive sector. Therefore, in this work, we consider maximum throughput prediction enhancing, for example, streaming or high-definition mapping applications. We discuss the entire \acrlong{ml} workflow highlighting less regarded aspects such as the detailed sampling procedures, the in-depth analysis of the dataset characteristics, the effects of splits in the provided results, and the data availability. Reliable \acrlong{ml} models need to face a lot of challenges during their lifecycle. We highlight how confidence can be built on \acrlong{ml} technologies by better understanding the underlying characteristics of the collected data. We discuss feature engineering and the effects of different splits for the training processes, showcasing that random splits might overestimate performance by more than twofold. Moreover, we investigate diverse sets of input features, where network information proved to be most effective, cutting the error by half. Part of our contribution is the validation of multiple \acrlong{ml} models within diverse scenarios. We also use \acrlong{xai} to show that \acrlong{ml} can learn underlying principles of wireless networks without being explicitly programmed. Our data is collected from a deployed network that was under full control of the measurement team and covered different vehicular scenarios and radio environments. 
\end{abstract}

%
\IEEEpeerreviewmaketitle

\newpage 
\section{Introduction}\label{sec::introduction}
\Glsfirst{ml} has a strong potential to overcome challenges arising in vehicular communication and networking, as presented in \cite{tang2020, tan2022machine}. Examples of such challenges are resource allocation \cite{ye2018deep, zhao2019deep} and \gls{qos} prediction \cite{kuelzer2021latency}. The \Gls{qos} prediction, in turn, is an enabler for a variety of use cases in the domain of vehicular communication, such as autonomous driving, platooning, cooperative maneuvering, tele-operated driving, and smart navigation. Some of these are presented in \cite{kuelzer2021ai4mobile, boban2018connected}.
Predictions can not only enhance the existing use cases and support new ones, but also be a stepping stone for a network to become proactive. Such predictions can be an integral part of a network that proactively reacts to sustain, for example, a specific \gls{kpi} target, like the maximum achievable throughput under high network load.

In this work, we focus on the prediction of \gls{qos} as a way of looking into the typical workflow for building and applying \gls{ml} algorithms. We start with data acquisition, moving on to data analysis and statistical characterization of the data, proceeding to feature engineering and different split strategies, and concluding by testing the performance of the model. 
Predicting \gls{qos} is a relatively complex task since it depends on several time-varying factors. It is especially complex in the case of vehicular communication where the radio conditions might change very drastically in a short amount of time \cite{torres2020qos}. 

Capturing datasets that reveal the complex inter-dependencies between network operations, radio environment and terminal behavior is a challenging task in itself. The captured data might often lack the required quality for \gls{ml}-based prediction, rendering it inappropriate. There are multiple factors that can reduce the quality of a collected dataset for training \gls{ml} models, with a few examples being dataset imbalance, coverage of radio regions with bounded dynamics or limited network states gathered. Moreover, acquiring different types of data for a prediction task comes with varying acquisition costs that are typically not discussed extensively.
From the theoretical perspective of statistical learning, a training dataset should often be \gls{iid} and drawn from the same probability distribution as the test dataset~\cite{shalev-shwartz2014understanding,mohri2018foundations}.
Radio terminals frequently experience drastic shifts in the data distribution~\cite{sarnelle2015} that usually degrade the performance of \gls{ml} models~\cite{redko2019advances}.

Meaningful comparisons between prior art proposals, though desirable, can be hard. There are multiple reasons for this, for example, the pre-processing steps are not well reported, or the used performance metrics differ.
Additionally, literature in the area of applying ML for wireless communications often reports the performance of the models without giving an in-depth analysis of the dataset’s underlying characteristics. This raises the question of how robust such models are to changes in different environments, where the dataset’s statistical properties might be different. 

Our results emerge from the data analysis of a dedicated measurement campaign with the goal of capturing several characteristics of the radio environment and enabling an in-depth discussion of \gls{ml} workflows.
The code for this analysis is based on customary functions from usual Python libraries for scientific computing
(e.g., numpy, pandas, scikit-learn...).
Our contributions include the following:
\begin{itemize}
\item    We study the prediction of maximum throughput prediction enhancing, for example, streaming or high-definition mapping applications.  

\item    We discuss the entire machine learning workflow highlighting less regarded, but important aspects, such as the detailed sampling procedures, the in-depth analysis of the dataset characteristics, and the effects of splits on performance. For example, we test against the data stationarity assumption and discuss methods to handle those for commercial networks. Many proposals in the literature avoid discussing such theoretical violations that many ML models require. 

\item    We put the characteristics of wireless environments in the foreground explaining what might hinder and benefit the adoption of ML approaches for new-generation wireless networks. 

\item    We further showcase why the lack of this type of information, missing often from the literature, can generate overconfidence in proposed ML solutions. 

\item    We investigate the effects of diverse sets of input features specifically those that are not typically available in the prior art.  

\item    We showcase that consumer devices and low sampling speeds can be enablers for the adoption of ML in commercial networks. 

\item    We also use explainable AI to demonstrate that machine learning can learn underlying principles of wireless networks without being explicitly programmed. 

\item    Our data is collected from a deployed network that avoids limitations of theoretical studies or simulation-based results. 
\end{itemize}
The remaining of the paper is structured as follows:
Section~\ref{sec::sota} presents related work in the area of \gls{pqos}, and the measurement campaign is described in Section~\ref{sec::measurement_campaign}. In Section \ref{sec::StatisticalProperties},
we discuss various radio environment properties in relation to \gls{ml} workflows, such as vehicle speed and stationarity. We continue the analysis on \gls{ml} workflows in Section~\ref{sec::MLforRadio} with a careful look into train/test data splitting and feature engineering. In Section~\ref{sec::ThroughputPrediction}, we apply a set of \gls{ml} algorithms to predict the maximum achievable throughput for diverse prediction models, communication directions, feature sets, splitting strategies, and prediction horizons. In Section~\ref{sec::interpretability} we look at the topic of \acrfull{xai}~\cite{8466590} showcasing what the models have learned in the absence of explicitly programmed rule sets. Finally, we draw conclusions in Section~\ref{sec::conclusion}.

\section{Related Work}\label{sec::sota}
The application of \gls{ml} for \gls{qos} prediction in vehicular networks has drawn attention recently. In the following, we present an overview of related works.
Measurement campaigns that focus on the feasibility of \gls{tod} are described in ~\cite{neumeier2019, gaber2020}. The authors of ~\cite{neumeier2019} also include a sensitivity analysis in their work and examine the impact of factors such as speed and distance to a serving cell.
When applying gathered data from measurements to \gls{qos} prediction in the current literature, a sampling interval in the order of seconds is often used (e.g., \cite{narayanan2020lumos5g, kulkarni2019deepchannel}). This does not reflect the fast-changing wireless channel associated with high-mobility vehicular communication scenarios~\cite{mecklenbrauker2011}, and there is an open question as to which degree sampling schemes need to be adjusted based on the vehicle speed. Some works already have compared measurement applications running on slow-sampling \gls{gnet} with \gls{minipc} analysis~\cite{lauridsen2016verification, alshamisi2018verifying} and propagation model tuning~\cite{enami2018}.
As vehicular communication is characterized by high mobility of the terminal, it is an open question whether \gls{gnet}'s low-cost transceiver chains might render collected data unsuitable for applying \gls{ml}. 

\par
Moreover, it is known that the radio environment has dynamics that can drastically change its statistical properties over a distance of a few meters \cite{ispas2010analysis, bernado2012validity}. Such changes in statistical properties of the data stream, where the \gls{ml} model needs to reformulate its decision boundaries, are being discussed in the literature as concept drift \cite{choudhary2017runtimeefficacy, 8926446, Schlimmer1986}. Detection and analysis of concept drift using state-of-the-art algorithms have been explained in \cite{smartcities4010021}. Extensive research is conducted on the management of concept drift, some of which are: using an ensemble learning approach~\cite{ensemble2021}, concept drift-aware federated averaging~\cite{Casado2022}, or model updating mechanisms~\cite{Zhang2016}.

In \cite{schmid2019comparison} and \cite{schmid2019deep} the authors present a measurement campaign and evaluate several \gls{ml} models regarding their performance for \gls{dll} throughput prediction. One limitation is that the authors measure in a public network and thus cannot consider features such as the cell load. The authors of~\cite{bo2018trust} consider the movement of the user and build a two-staged \gls{ml} model to predict the \gls{tcp} throughput. In the first step, the movement pattern is identified. Based on this, a \gls{lstm}, trained with corresponding data, is selected for the throughput prediction. Some preliminary results for \gls{ul} and \gls{dll} throughput prediction using Random Forest and employing the dataset used for this paper are presented in \cite{palaios2021effect}. \Gls{qos} prediction in a 5G non-standalone network is examined in \cite{sliwa2021machine}. Besides throughput, \gls{ml} is also used to predict latency~\cite{khatouni2019machine, kuelzer2021latency, mankowitz2011mobile} and handovers~\cite{feltrin2018machine}. Recently, several datasets intended for \gls{ml}-based studies have been made publicly available \cite{burmeister2021measuring, farthofer2022an} that typically do not provide cell-related information. Moreover, those works focus on introducing the datasets instead of applying \gls{ml} methods.

An inherent limitation of \gls{ml} algorithms, especially as they get more complex, is that their decision-making is difficult to understand, rendering them into black boxes. Recently, the field of \gls{xai}~\cite{8466590} has gained traction, to ensure ML models perform as desired, not only in test environments but also in target applications. With \gls{xai}, models can be judged in the context of human domain knowledge, e.g. by investigating the learned feature importance.

We extend prior art by providing detailed insights into the aspects mentioned above. Our dedicated measurement campaign, on a fully controlled private network, allowed us to perform a deep analysis and enable an in-depth discussion. We could control which devices could connect to the network, we could adjust the total network interference, and generate diverse traffic dynamics. We consider different radio environments, device types, sampling frequencies, measurement scenarios, and prediction horizons. We present the complete workflow towards building a \gls{pqos} model, predicting the maximum achievable throughput. We discuss several design decisions, such as the train-test split, various feature groups, and different models, and show the influence on the achieved performance. In the end, we discuss the issue of trustworthiness.

\begin{figure}[h!t]
\begin{center}
\includegraphics[width=.7\columnwidth]{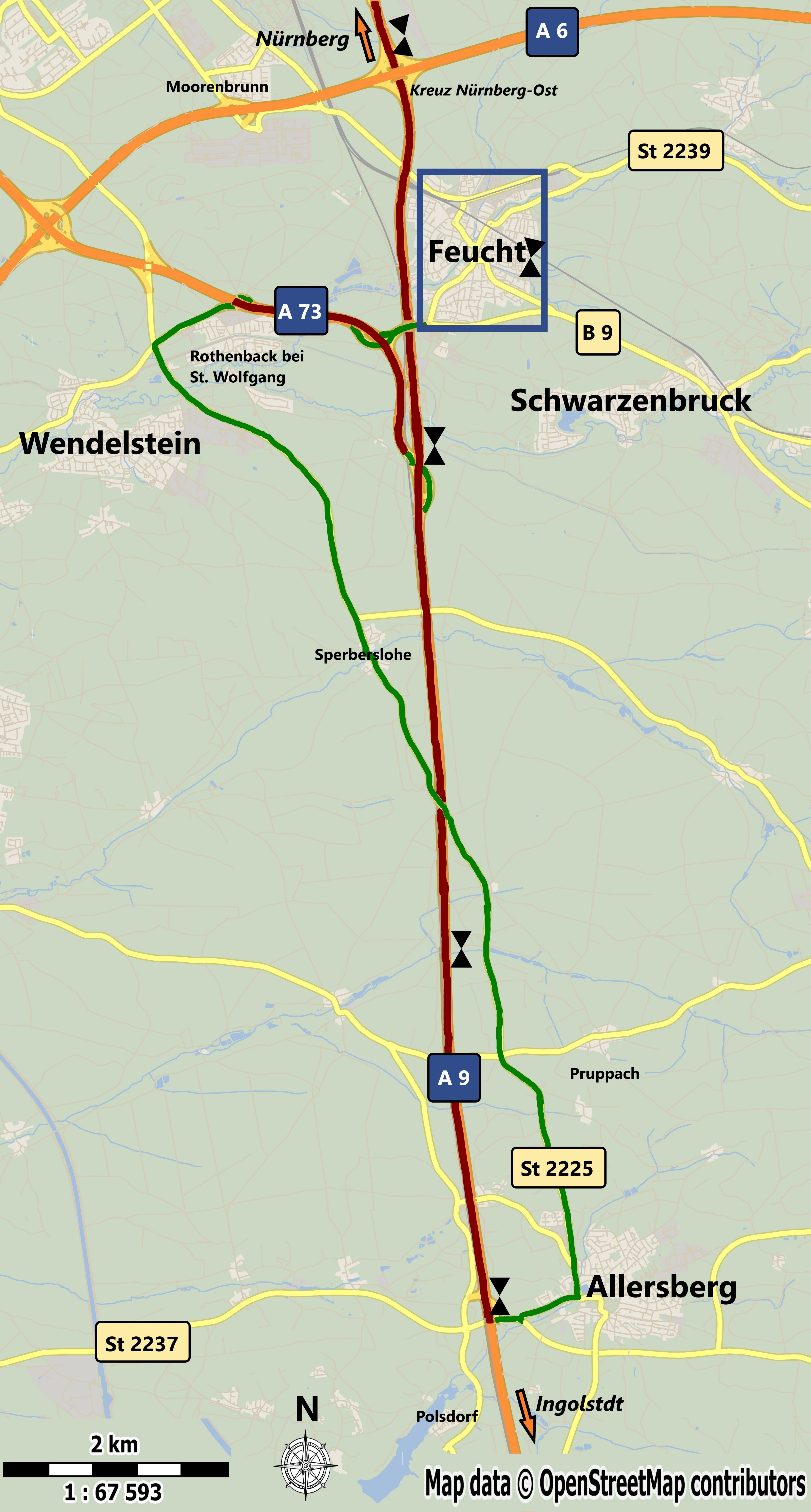}
\end{center}
\vspace*{-3mm}
\caption{The map of the areas where the measurements took place. The A9 highway is highlighted in red, while the rural and side street areas are marked in green (not all rural streets are plotted to reduce clutter). The blue area is the suburban city of Feucht.}
\label{fig::Map}
\vspace*{-2mm}
\end{figure}

\section{Measurement Campaign}\label{sec::measurement_campaign}
Within the project AI4Mobile, an extensive measurement campaign was performed in the 5G-ConnectedMobility test field\footnote{https://www.ericsson.com/en/cases/2020/5g-connectedmobility}. In total, more than 3000\,km were driven by four vehicles for one week to perform measurements from the \gls{lte} network. A detailed description of the measurement campaign is available in~\cite{palaios2021measurements}. Since we had control over the network infrastructure, we were able to measure and collect data from all parts of the network. We were also able to define custom scenarios and set parameters accordingly. 
Subsequently, we briefly describe the setup and highlight some scenarios.

\subsection{Measurement Setup}

During the measurement campaign, four vehicles were equipped with several \glspl{ue}. In each vehicle, an identical \gls{minipc} was used for measurements with high granularity. The \gls{minipc} is a Linux-based PC including a cellular modem connected to a 2x2 MIMO car antenna, which is placed on the roof of the respective car. Additionally, a \gls{GPS} receiver was placed on the roof and connected to the \gls{minipc}, providing accurate location and time synchronization via a \gls{pps} signal. To capture the data from the cellular modem, we used the application MobileInsight \cite{li2017mobileinsight}.

Each vehicle was also equipped with several \gls{gnet} phones as pure data generators and measurement devices with a low sampling frequency for a comparison to the \gls{minipc}'s measurements.
The \gls{gnet} uses the application G-Net Track Pro \cite{gyokovsolutions_2016} to capture radio measurements via the Android API. All devices used the application Iperf \cite{iperf} to exchange data with a local server connected to the network. In addition to the \glspl{ue} (\gls{gnet} and \gls{minipc}), we also collected data from the base stations and the core network, which were also time-synchronized. An overview of all captured data is given in Table~\ref{tab::CapturedData}.

The network is deployed in the south of Germany, close to Nuremberg. It consists of 10 \gls{fdd}-\gls{lte} cells with 10\,MHz of bandwidth at a carrier frequency of 700\,MHz, with typical radio tower panel antennas (horizontal half-power beam width of 65$^{\circ}$). Since it is a test network, our devices were the only devices connected to the network. In Fig. \ref{fig::Map}, the area of the test field is shown, including the position of the distinct cells. While the cells were positioned to provide excellent connectivity along the highway and in the suburban city of Feucht, part of the rural area next to the highway was also covered, which allowed us to capture several radio environments.

\subsection{The Conducted Studies}
The purpose of the measurements was to thoroughly examine numerous parameters through a variety of real-time scenarios in our captured dataset. This involved analyzing different driving patterns at different speeds, stationary measurements, recording different radio environments, and generating varied data traffic to create contrasting load scenarios. Furthermore, we utilized various protocols and generated traffic in both the \gls{ul} and \gls{dll} to the server and to other vehicles. Additional information on the parameters captured in the measurement campaign can be found in~\cite{palaios2021measurements}.

\begin{figure}[t]
\begin{center}
\includegraphics[width=1.0\columnwidth]{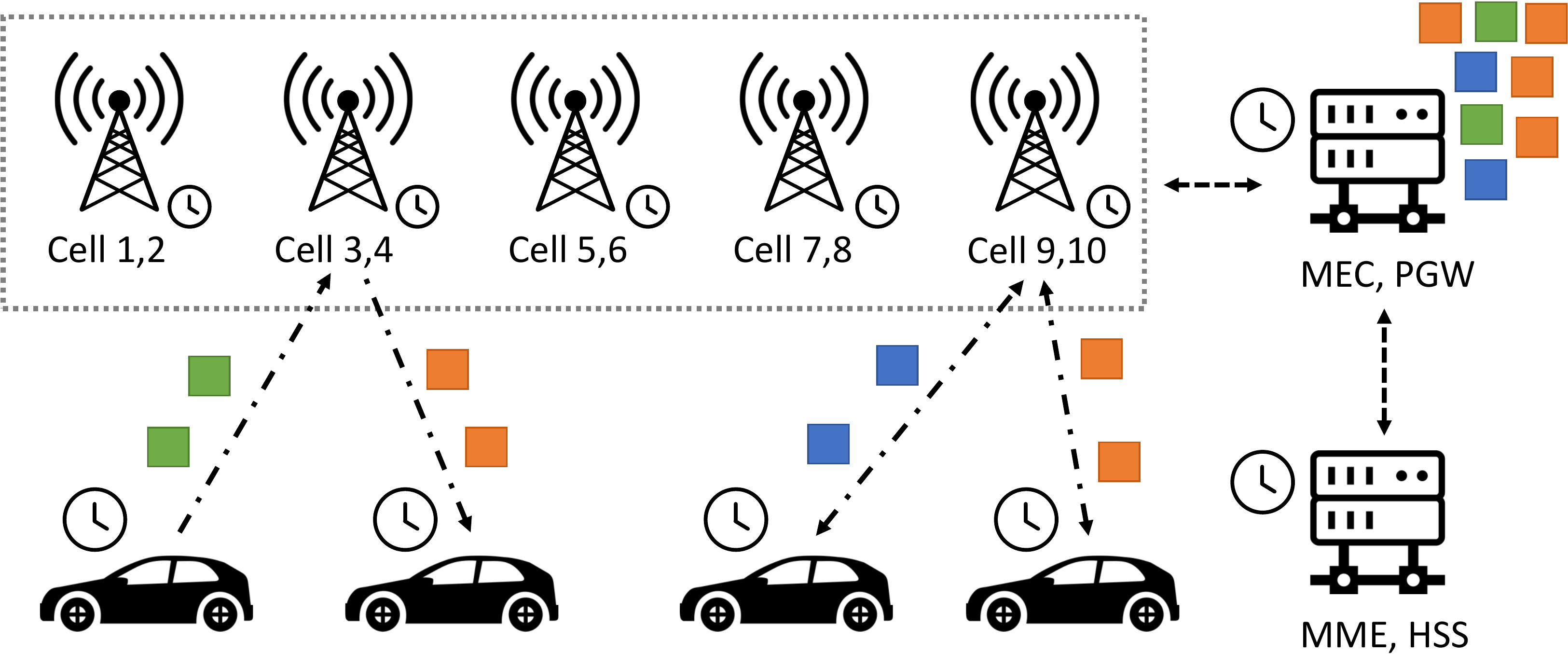}
\end{center}
\vspace*{-3mm}
\caption{Simplified view of the used measurement methodology. All clocks are synchronized and all generated data is marked accordingly and captured at separate entities. Figure adapted from~\cite{palaios2021measurements}.}
\label{fig::MeasurementMethodology}
\vspace*{-2mm}
\end{figure}

\begin{table*}[t]
\centering
\setlength{\tabcolsep}{4pt}
\caption{Overview of captured data from all network nodes.}
\label{tab::CapturedData}
\scriptsize
\begin{tabularx}{\textwidth}{lllll}
\toprule
\textbf{Network Node} & \textbf{Information Type} & \textbf{Sampling Interval} & \textbf{Features (Examples)} & \textbf{QoS Measurements}\\
\cline{1-5}\Tstrut
\multirow{7}{*} {\textbf{DME}}  & \Acrfull{phy} & \SI{10}{\milli\second} & RSRP, RSRQ, CQI & \\
            & \Acrfull{mac} & \SI{40}{\milli\second} & \Acrfull{bsr}
            &\\
            & \Acrfull{pdcp} & \SI{0.5}{\second} & PDCP UL \& DL statistics &\\
            & \Acrfull{rlc} & \SI{45}{\milli\second} & RLC transmissions \& receptions &\\
            & \Acrfull{rrc} & Event-based & OTA packet log &\\
            & Built-in sensors (e.g., GPS) & \SI{1}{\second} & Longitude, latitude, speed &\\
            & Packet capturing tool (Tcpdump) & Traffic-dependent & & DL throughput \& latency\\
\cline{1-5}\Tstrut
\multirow{2}{*} {\textbf{CE}} & \Acrfull{phy} & \SI{1}{\second} & RSRP, RSRQ &\\
            & Measurement tool (Iperf) & \SI{1}{\second} & & DL throughput\\
\cline{1-5}\Tstrut
\multirow{2}{*} {\textbf{Base station}} & Standard operator averages & \SI{15}{\minute} & Interference, cell load &\\
            & Reconstructed real-time information & \SI{1}{\milli\second} & Interference, cell load &\\
\cline{1-5}\Tstrut
\multirow{2}{*} {\textbf{Core}} & \Acrfull{mme}
Traces & Event-based & Handover &\\
            & Packet capturing tool (Tcpdump) & Traffic-dependent & & UL throughput \& latency\\
\cline{1-5}\Tstrut
\multirow{2}{*} {\textbf{Database}} & Weather & 1 hour & Precipitation &\\
            & Traffic APIs & 5 min & Traffic flow &\\
\bottomrule
\end{tabularx}
\end{table*}

\subsection{Data Preprocessing}
For this study, only \gls{udp} measurements from the dataset were considered since the protocol properties of \gls{tcp} and \gls{udp} differ. Therefore the E2E throughput could not be limited by the transport layer. 
We used the data, where the devices were requesting maximum throughput.

Moreover, we filtered out stationary measurements, i.e., when vehicles were parked for some time. The stationary measurements contain similar radio properties for the majority of the samples, which could result in overly optimistic prediction results.

\section{Properties of the Radio Environment}\label{sec::StatisticalProperties}
In this section, we study properties of the radio environment that might influence \gls{ml} algorithms. Radio environments have very diverse characteristics that depend on multiple factors. Some of these factors are environment-based, for example, radio propagation can be drastically different in a rural area and in a crowded city center~\cite{dyspanLondon}, as the type and density of buildings influence the radio propagation. Another factor is, for example, the specific network deployment (e.g. antenna heights, frequency layers and the number of cells in an area). All these factors contribute to very unique properties in the collected data as UEs move. We first discuss the data collection procedures and how these might affect the quality of the collected dataset. We continue looking at the statistical properties of the radio environment. As many \gls{ml} algorithms rely on multiple statistical assumptions~\cite{sayed2012learning}, we try to highlight some of those, as they might need to be better understood by the research community that applies \gls{ml} on wireless networks.

\subsection{Sampling the Radio Environment}\label{sec::SamplingRadio}

High sampling intervals of the radio environment come at a cost, either in terms of acquiring more capable hardware, or in terms of higher power consumption or in terms of signaling, as more data need to be transferred. We specifically study the effects of different sampling intervals on the characteristics of the collected dataset. We also study the effects of the vehicular speed in the characteristics of the collected dataset to see if higher velocities call for faster sampling intervals, for sustaining specific quality characteristics in the collected dataset.
We focus on \gls{lte} signal values for the analysis, i.e., \gls{rsrp} and \gls{rssi}, since these are most affected by the vehicle's speed and influence the target throughput, in turn, as we show in Section~\ref{sec::ThroughputPrediction} and in Section~\ref{sec::interpretability}

\begin{figure*}
\begin{subfigure}{.24\textwidth}
    \centering
    \includegraphics[width=\columnwidth]{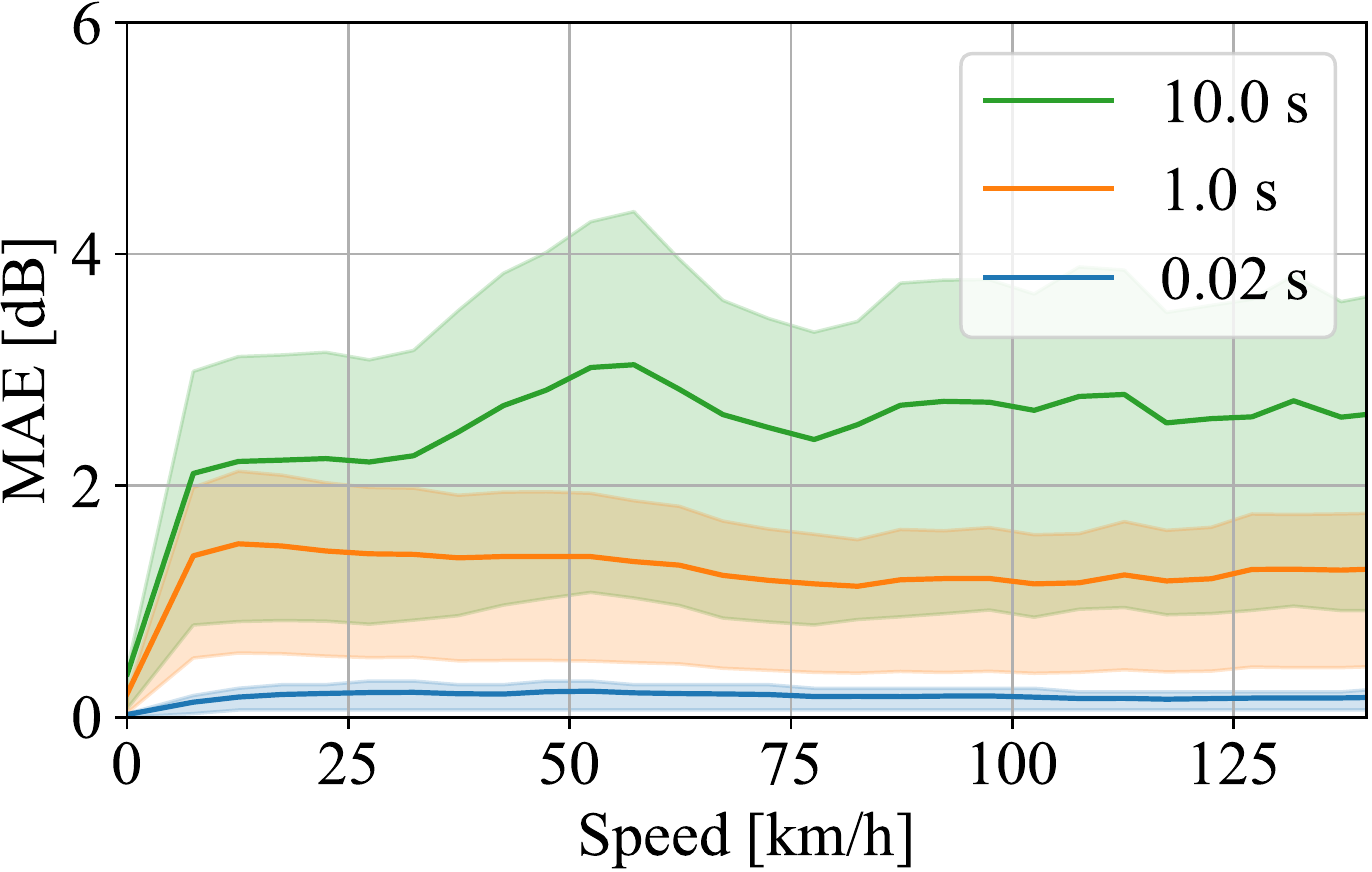}
    \caption{\gls{rsrp} (\gls{minipc}/10\,ms)}
    \label{fig::speed_error_rsrp_minipc}
\end{subfigure}
\begin{subfigure}{.24\textwidth}
    \centering
    \includegraphics[width=\columnwidth]{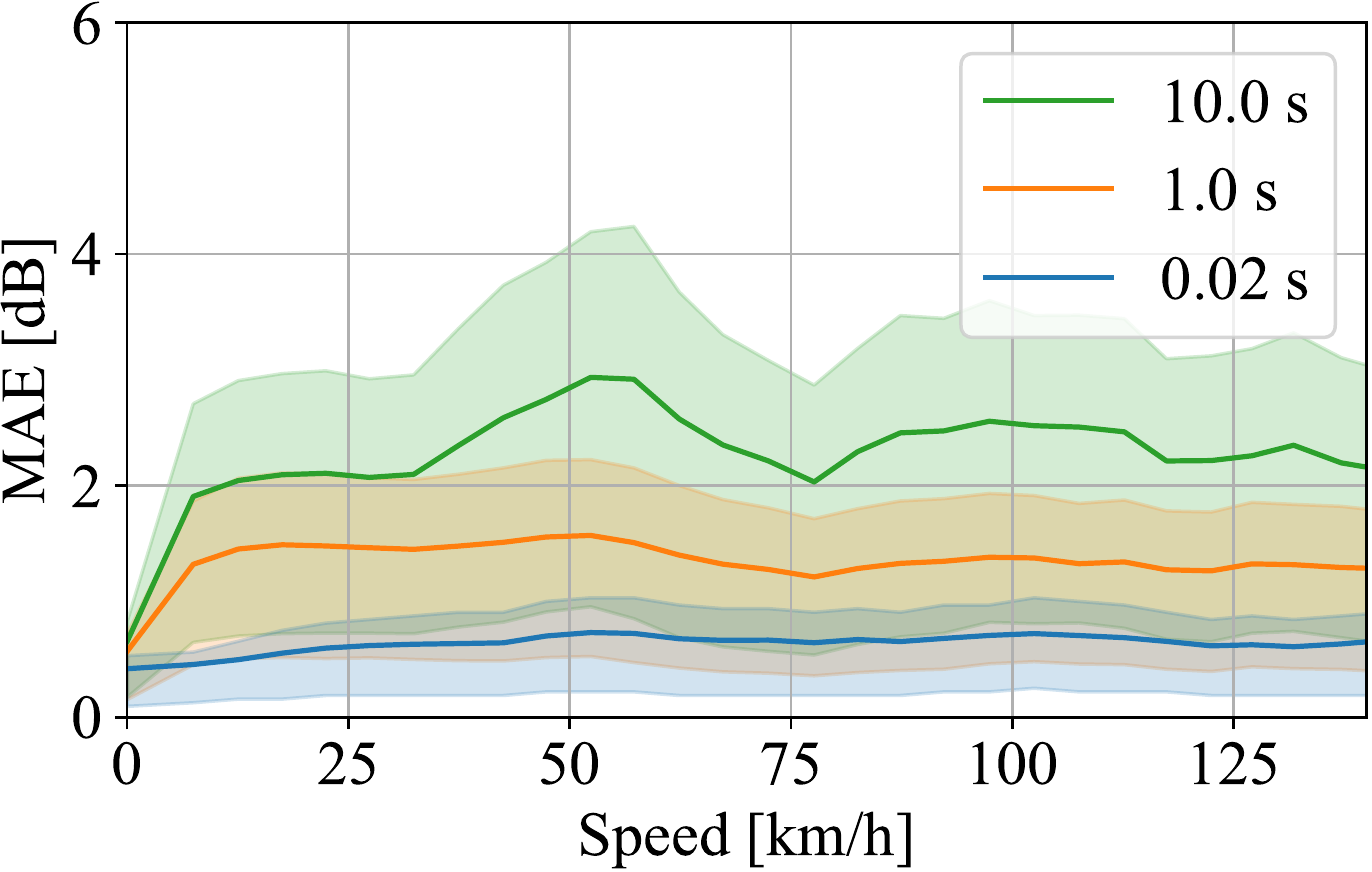}
    \caption{\gls{rssi} (\gls{minipc}/10\,ms)}
    \label{fig::speed_error_rssi_minipc}
\end{subfigure}
\begin{subfigure}{.24\textwidth}
    \centering
    \includegraphics[width=\columnwidth]{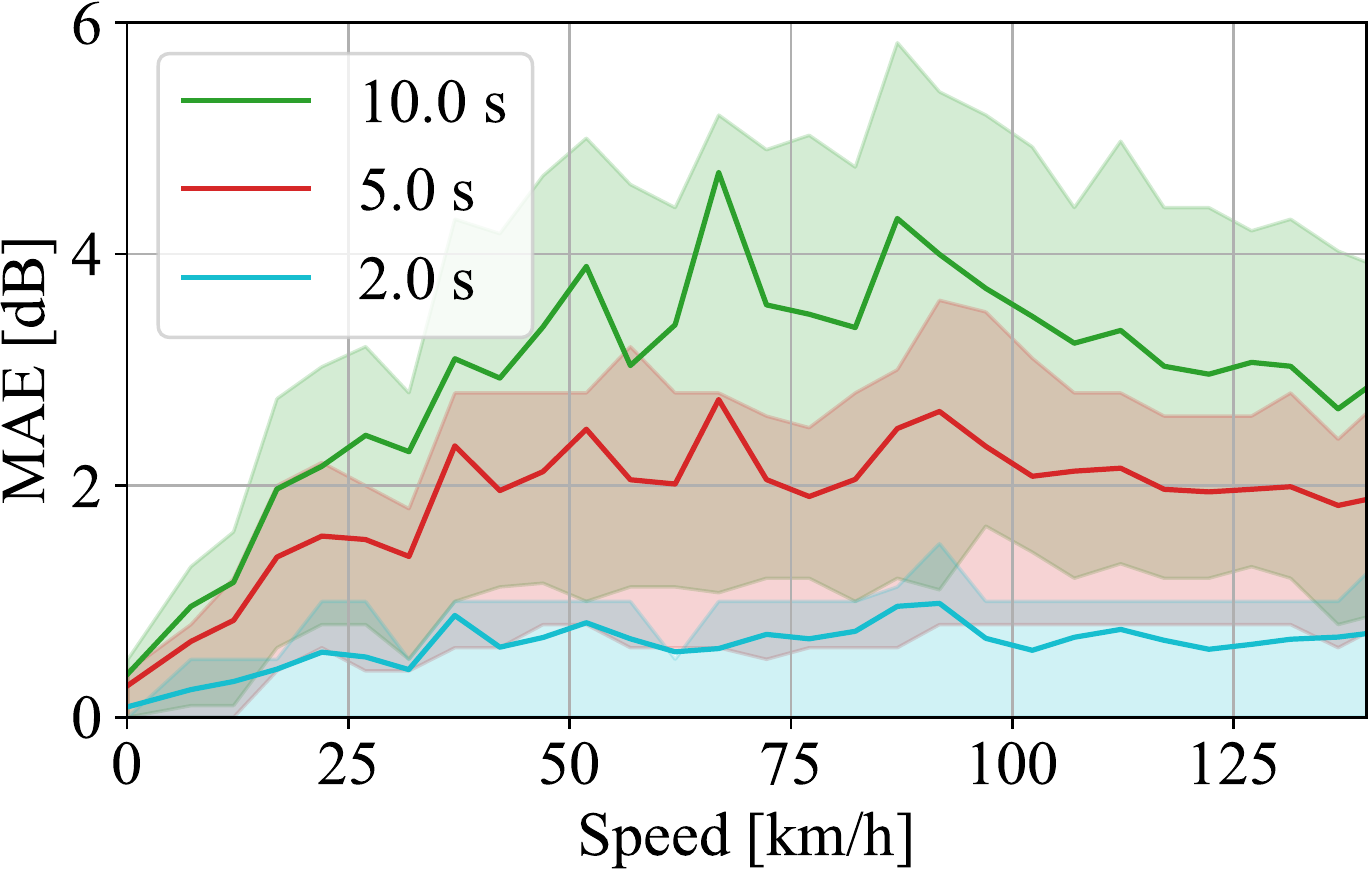}
    \caption{\gls{rsrp} (\gls{gnet}/1\,s)}
    \label{fig::speed_error_rsrp_gnet}
\end{subfigure}
\begin{subfigure}{.24\textwidth}
    \centering
    \includegraphics[width=\columnwidth]{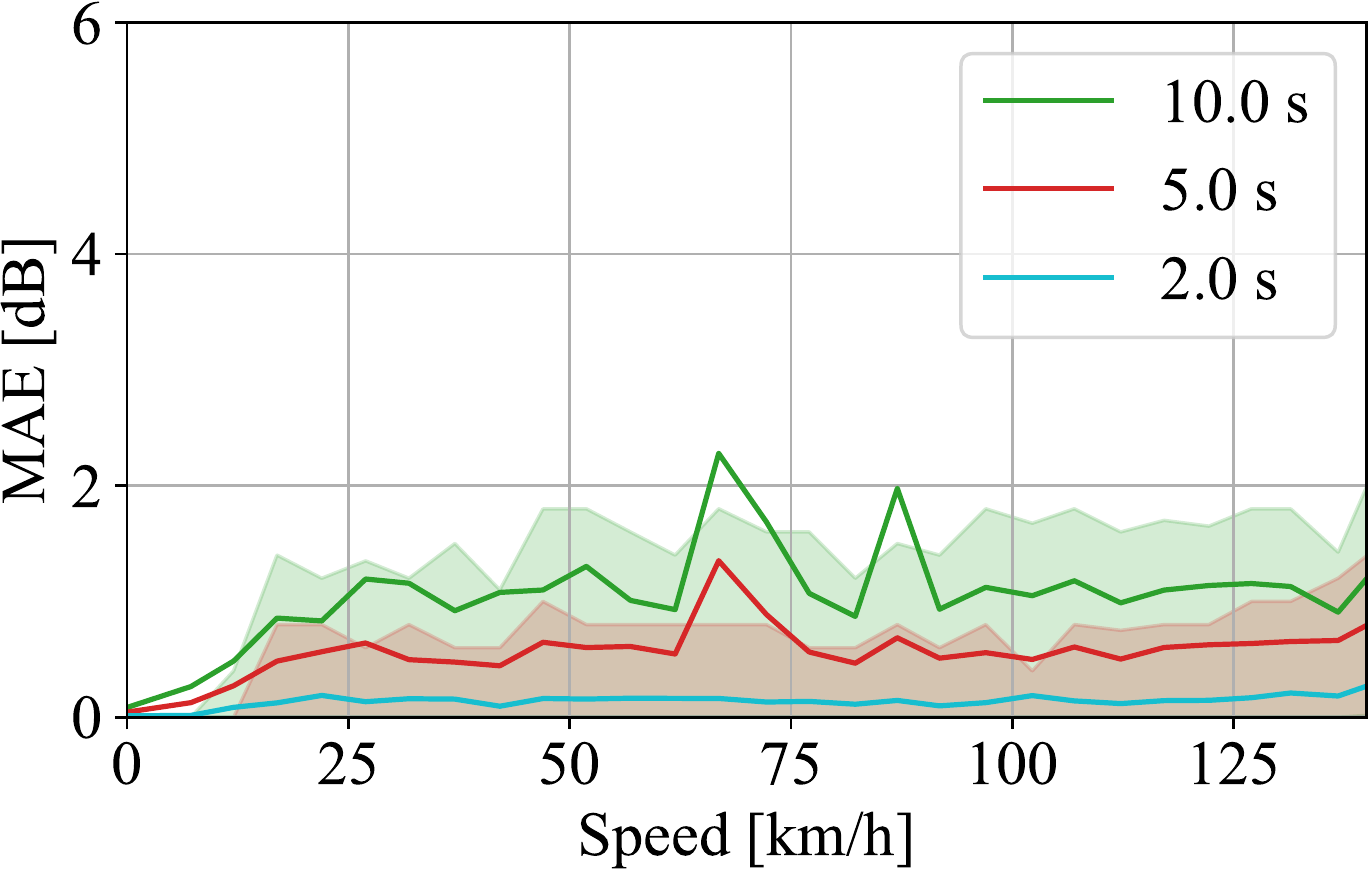}
    \caption{\gls{rssi} (\gls{gnet}/1\,ms)}
    \label{fig::speed_error_rssi_gnet}
\end{subfigure}
\caption{Relation between speed and mean absolute error (MAE) for different sampling intervals (shaded areas indicate the interquartile range).}
\label{fig::speed_error}
\end{figure*}
To quantify the sensitivity of the collected measurements on these two parameters, we need to introduce the \gls{de} at the $k$-th time instance as
\begin{equation}
\label{eq::DE}
    RE(k) = x_{o}(k) - x_{rs}(k) \; , \; 0 \le k < K \; , 
\end{equation}
where $x_{o}(k)$ represents the original signal of length $K$ at a sampling interval of \SI{10}{ms} for the \gls{minipc} and \SI{1}{s} for the \gls{gnet}. The corresponding resampled value for a lower sampling interval is $x_{rs}(k)$. To determine $x_{rs}(k)$, the following procedure is applied: first, the original dataset is downsampled by averaging to a target rate to obtain $x_{ds}\left(k'\right)$:
\begin{equation}
\label{eq::ds}
    x_{ds}\left(k'\right)=\frac{1}{M}\sum_{m=0}^{M-1}{x_o\left(m+k'M\right)} \; , \; 0 \le k' < K/M \; , 
\end{equation}
where $M$ accounts for the downsampling ratio for a certain pair of original and target sampling interval, e.g., $M=10$ for a target rate of \SI{10}{s} when considering the \gls{gnet}'s original sampling interval of \SI{1}{s}. Secondly, the downsampled dataset are then upsampled by the forward filling method, which can be expressed by $M$-integer division on the original index $k$:
\begin{equation}
\label{eq::us}
    x_{rs}\left(k\right)=x_{ds}\left(\left\lfloor k/M \right\rfloor\right)=x_{ds}\left(\frac{k-(k\mod{M})}{M}\right)\;.
\end{equation}
Finally, Equations (\ref{eq::DE}), (\ref{eq::ds}) and (\ref{eq::us}) can be combined into
\begin{equation}
    RE(k) = \frac{1}{M}\sum_{m=0}^{M-1}{x_o\left(k\right)-x_o\left(m+k-(k\mod{M})\right)}
    \;.
\end{equation}

For each sample at the $k$-th time instance, the vehicle speed $v(k)$ is also known, which allows to explore the relation between the mobility of the terminal and the resulting \gls{de}. Therefore, the speed range from 0 to \SI{140}{\kmh} is split in intervals of \SI{5}{\kmh}. For each interval with a lower speed boundary $v'$, the mean absolute \glsdesc{de} $\overline{RE}_{v'}$ is calculated according to Eq.~\eqref{eq::mean_DE}. 

\begin{equation}
\label{eq::mean_DE}
    \overline{RE}_{v'} = \frac{1}{K} \sum_{v' \leq v(k) < v'+5\,km/h} |RE(k)|
\end{equation}

The results are shown for \gls{rsrp} and \gls{rssi} and measurements from the \gls{minipc} and \gls{gnet} in Fig.~\ref{fig::speed_error}. The shaded areas indicate the central half of the distribution of the errors.

Focusing on the effect of different sampling intervals for a \gls{minipc} dataset (Figs.~\ref{fig::speed_error_rsrp_minipc} and \ref{fig::speed_error_rssi_minipc}), one can see that the \gls{de} increases with a larger sampling interval as expected. Interestingly, this increase seems to be below \SI{2}{dB} for a sampling interval of \SI{1}{s} and between 2 and \SI{3}{dB} for a sampling interval of \SI{10}{s}, which might still be acceptable for certain \gls{ml} applications.

The mobility of the terminal seems to have some effect on the quality of the collected dataset. When the terminal is stationary, the \gls{de} is close to 0. Both \gls{rsrp} and \gls{rssi} are characterized by a steep rise of the error once the terminal starts to move at an approximate speed of \SI{10}{\kmh}. Above that speed, the error remains stable. Of particular interest is a sampling interval of \SI{1}{s}, as it shows approximately the error introduced by employing a \gls{gnet} instead of a \gls{minipc}, which collects data at a sampling interval of \SI{10}{ms}. The mean \gls{de} for \gls{rsrp} and \gls{rssi} are well below 2\,dB, which suggests that also \gls{gnet} samples may enable an accurate \gls{qos} prediction. Only the sampling interval of \SI{10}{s} shows larger variations across different speeds. Here we notice that the variance of the error is drastically increased for the lower sampling interval. 

We also executed the same experiments with a dataset from the less granular \gls{gnet}. We received similar results for the \gls{minipc} dataset as shown in Fig. ~\ref{fig::speed_error_rsrp_gnet} and \ref{fig::speed_error_rssi_gnet}, i.e., a rise of the error in the beginning, followed by a saturation of the \gls{de}. We empathize that a direct comparison of the resulting \gls{de} between the two different devices should be avoided, as the original sampling intervals are larger and thus the number of samples that are considered for each downsampling interval is smaller. This, combined with the simpler transceiver chain of the \gls{gnet}, is a potential explanation for why these results are less smooth.

The results indicate that the vehicle speed does not have a noticeable impact on the absolute error for different sampling intervals, except for vehicles moving at very low speeds. The observed effects can possibly be attributed to the Doppler effect. However, no channel measurements were carried out during the measurement campaign and thus the causes cannot be conclusively assessed on the basis of our data set. The relatively constant error for slow sampling at different speeds can further simplify the sampling procedures without the need to introduce adaptive sampling schemes. Also, lower sampling speeds can still provide useful input to \gls{ml} models as the reconstruction error remains low. This type of result is necessary for optimizing data collection procedures for future \gls{ml} applications.
In the following sections, we use downsampled data from the \gls{minipc} with a sampling interval of 1s that also enables a comparison with the \gls{gnet}.
The resulting dataset sizes for DL and UL are 18846 and 19535 samples, respectively.
\subsection{Data Stationarity}\label{sec::DataStationarity}
Many theoretical results in \Gls{ml} literature are based on the assumption that the
available dataset is \gls{iid}~\cite{shalev-shwartz2014understanding,mohri2018foundations,redko2019advances}.
This assumption can be violated in several ways, for instance, if
i) samples are correlated in time (i.e., they are not independent), or
ii) the underlying time series is non-stationary (i.e., not identically distributed).

There are both parametric and non-parametric hypothesis tests available in the literature to study time stationarity.
Parametric tests are often more powerful, but one needs to assure that the assumptions they
require (i.e., a specific well-known underlying distribution) are met by the tested data.
As such, we have opted to use the parametric
\gls{adf} test~\cite{ng1995unit}, which searches a unit root in the tested time series
as its null hypothesis,
while the alternative hypothesis is stationarity. In other words, a low $p$-value
(for our analysis we have considered $p<0.05$) indicates strong evidence for stationarity
in the data.
As a parametric test, \gls{adf} requires a suitable choice of the maximum lag
of the assumed autoregressive model.
Since the lag order is unknown for our data, we estimate it according to the \gls{aic}~\cite{ng1995unit}.

\begin{figure}
     \centering
     \begin{subfigure}[b]{0.8\linewidth}
         \centering
         \def\svgwidth{\linewidth}
         \scriptsize
          \includegraphics[width=\linewidth]{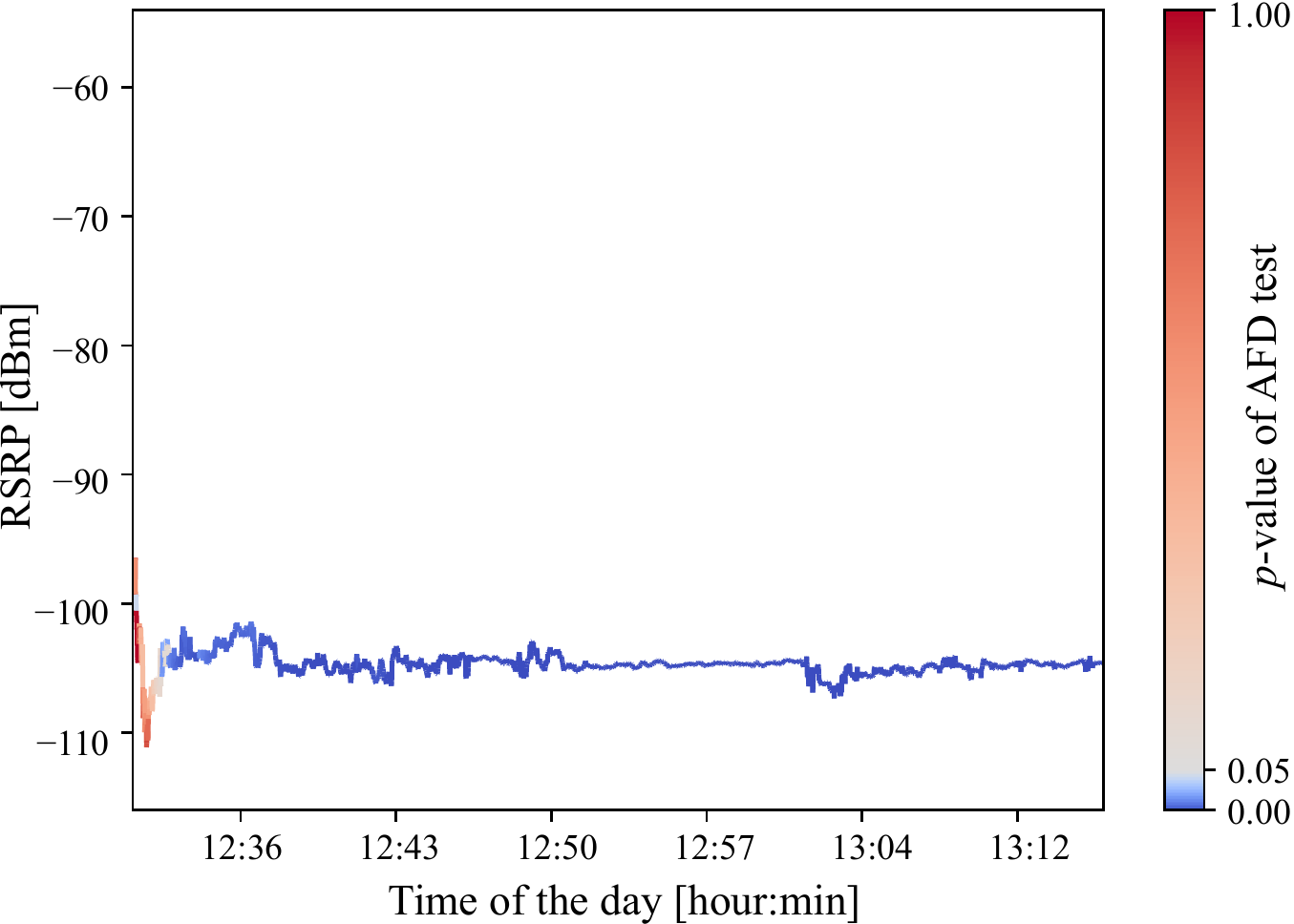}
         \caption{Low-variance scenario}
         \label{fig:lovar}
     \end{subfigure}
     \hfill
     \begin{subfigure}[b]{0.8\linewidth}
         \centering
         \def\svgwidth{\linewidth}
         \scriptsize
         \includegraphics[width=\linewidth]{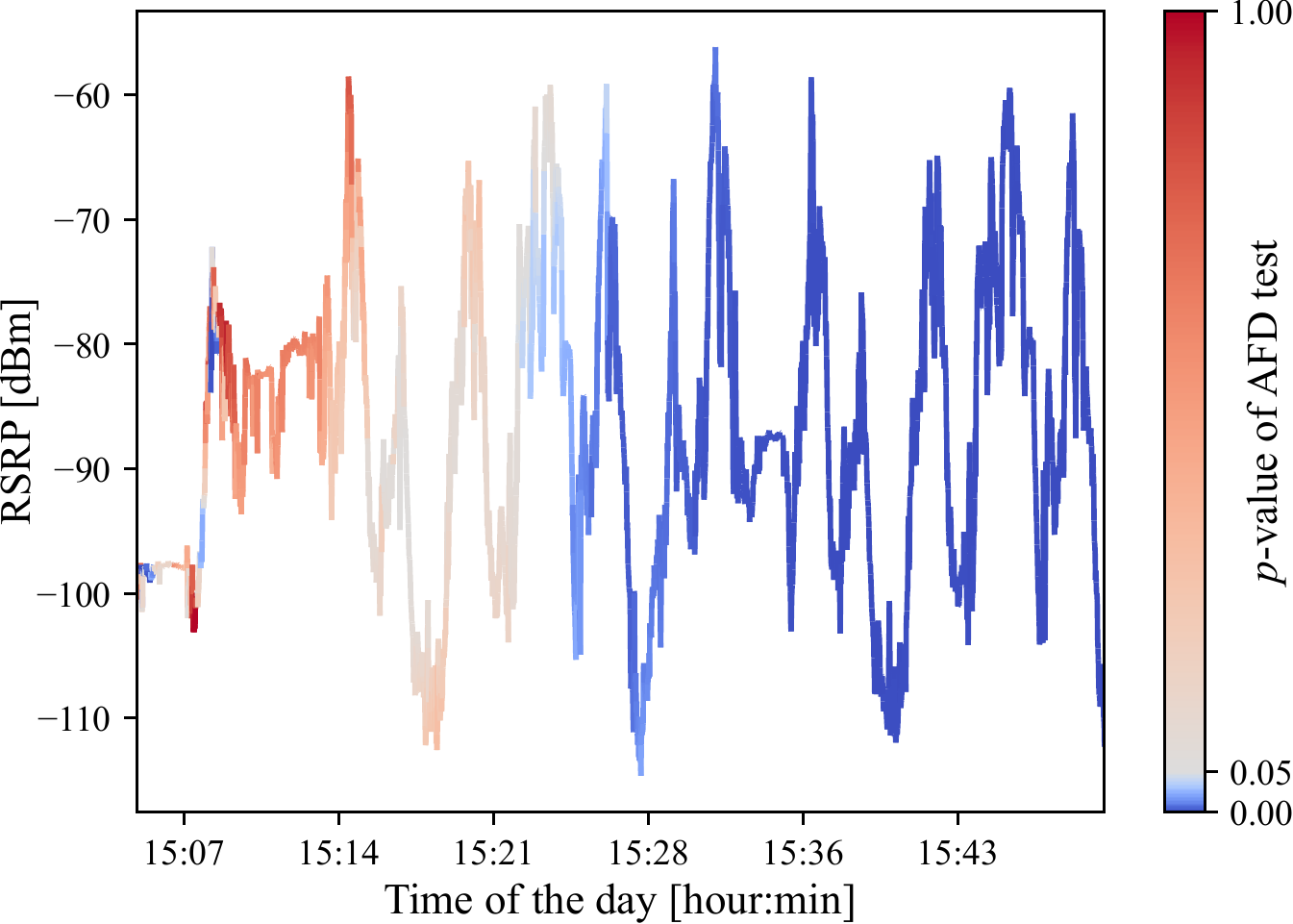}
         \caption{High-variance scenario}
         \label{fig:hivar}
     \end{subfigure}
     \hfill
     \begin{subfigure}[b]{0.8\linewidth}
         \centering
         \def\svgwidth{\linewidth}
         \scriptsize
         \includegraphics[width=\linewidth]{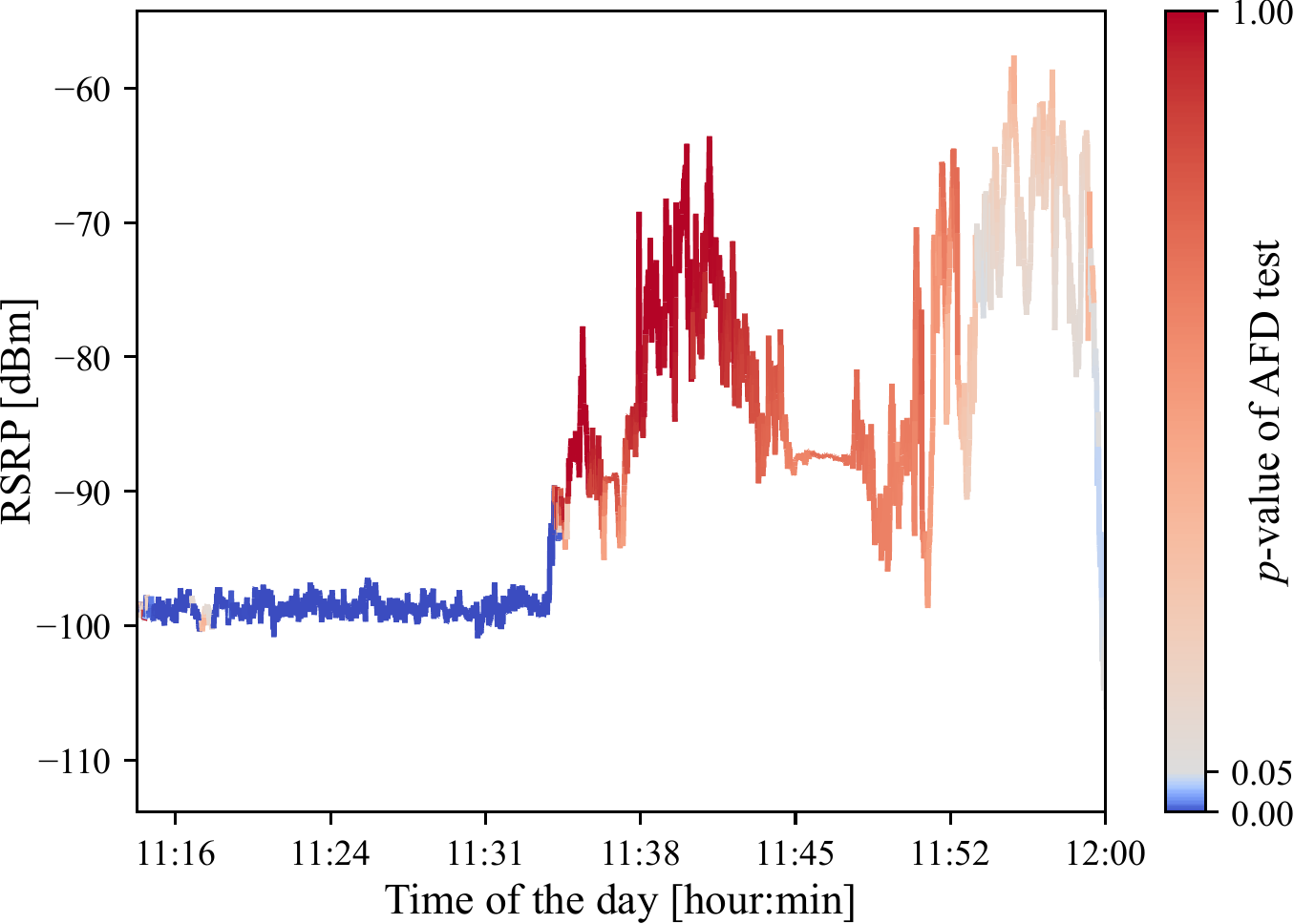}
         \caption{Scenario shift}
         \label{fig:scenshift}
     \end{subfigure}
        \caption{Stationarity analysis of the data using the \gls{adf} test.}
        \label{fig:stationarity}
\end{figure}

We have run the \gls{adf} test for all \gls{phy} features from all measurements (c.f.~\Cref{tab::CapturedData}, first row) in an accumulated manner.
That is, for every analyzed time series of size $T$, we construct $T$ sub-series each one consisting of all
samples from 0 to $t\in\left\lbrace 0,1,\ldots,T-1\right\rbrace$.
We then run the \gls{adf} test against every sub-series and concatenate the output
$p$-values into a vector of size $T$. Fig.~\ref{fig:stationarity}
shows the resulting accumulated \gls{adf} test
as a color code on top of some time series.

The expected behavior is that the accumulated \gls{adf} test tends to $p\ll0.05$ as the
size of the evaluated time sub-series increases. The number of samples needed for this depends
on some statistics, such as the variance (c.f. Fig. \ref{fig:lovar} \& \ref{fig:hivar}),
but it rarely exceeds 15 minutes for our data.

This time stationarity assumption is violated for measurements with an appreciable scenario shift, e.g., those where vehicles
remain idle for some time and drive away afterward (Fig.~\ref{fig:scenshift}).
This shows that stationarity cannot always be assumed, even though it might hold true
for some datasets under a long-enough measurement time. For the measurement scenarios, we opted for a duration of 40 to 60 minutes while driving to alleviate the problem of non-stationarities.  

In summary, we acknowledge the presence of non-stationary portions in our data and its possible negative impact
on the performance  of \gls{ml} algorithms. 

\subsection{Radio Environment Correlations}\label{sec::DataCorrelation}
\begin{figure}[htbp]
\centering
    \includegraphics[width=\wFig, height=\hFig]{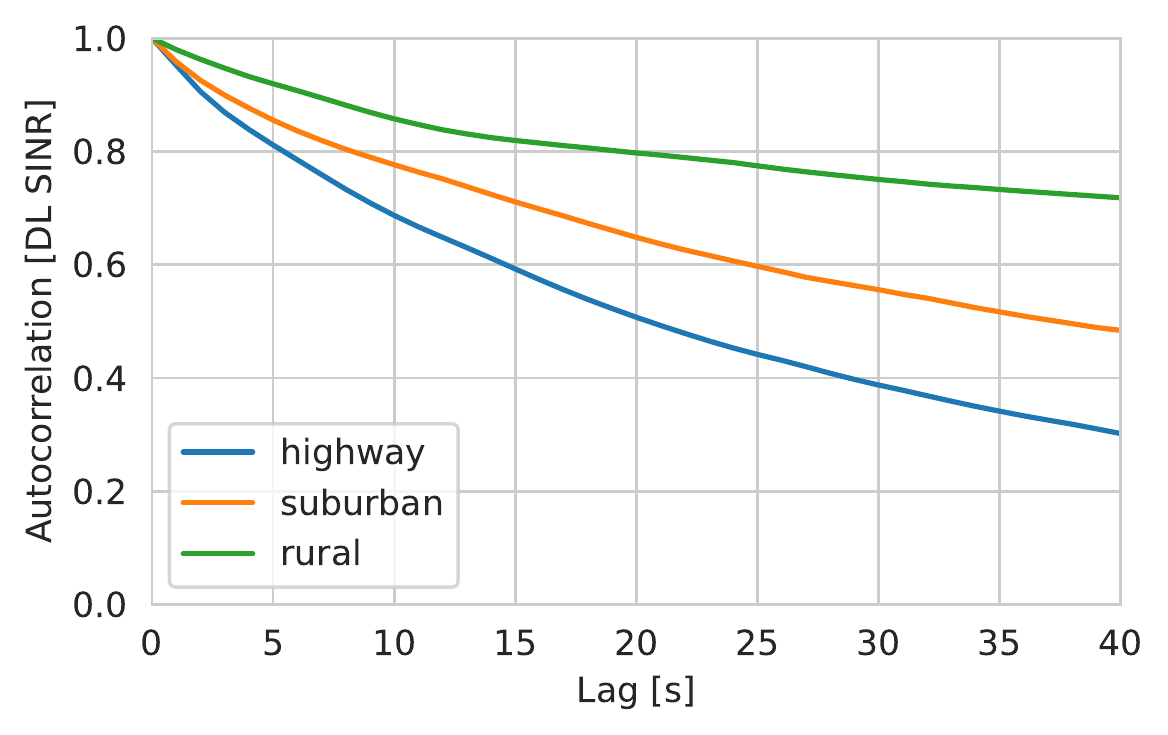}
    \caption{Autocorrelation of the SINR for excerpts of different DL measurement runs.}
    \label{fig::SINR_Autocorr}
\end{figure}

Additionally, we look at an important characteristic of the radio environment, which is the correlation in time and space~\cite{Perpinias,timecorrelations}. In Fig.~\ref{fig::SINR_Autocorr}, we show an example of the autocorrelation function of the \gls{sinr} for three distinct radio environments. We see that the radio environment tends to be highly correlated in time for all three distinct radio environments. Some higher correlations are noted for the rural environment, with the highway having the weakest of the three. This is probably explained by the faster movement of the vehicles.

In~\cite{cv_correlated}, the authors discuss that if correlations are not handled properly, there is a higher risk of reporting overly optimistic results. The reason is that the \gls{ml} model does not learn the underlying relations between input and output clearly, and instead learns the existing dependencies of a dataset, which in our case, are the reported correlations.

\section{Machine Learning for Wireless Networks}\label{sec::MLforRadio}
In this section, we discuss important aspects that the \gls{ml} engineer needs to understand before applying \gls{ml} to radio data. 
We discuss the notion of concept drift, different strategies for splitting the data into training and test datasets, and feature engineering. 

\subsection{Concept Drift}\label{sec::DataShift}
In this section, we introduce the idea of concept drift as a tool that can mitigate some of the effects of non-stationarity.
 \gls{ml} models traditionally need a formal and precise definition of the problem to represent the decision boundaries. However, non-stationarity in datasets implies that the decision boundaries, i.e., concepts, have a higher likelihood to be relearned to accommodate variations in the underlying data distributions \cite{widmer1996learning} \cite{tsymbal2004problem}. This re-learning or concept drifts must be detected to ensure that the \gls{ml} models remain accurate. Given that a sample instance $x$ belongs to a class $\omega$ ($x \in \omega$), the Bayes posterior probability is:
\begin{equation}
    \label{eq::concept_drift}
    P(\omega \mid x)=\frac{P(\omega)P(x \mid \omega)}{P(x)}
\end{equation}
This indicates that the drift may occur when:
\begin{enumerate}[label=(\roman*)]
    \item Class definition $P(\omega \mid x)$ changes while $P(x)$ remains the same, i.e., real concept drift.
    \item Virtual drift: Input distribution $P(x)$ changes while $P(\omega \mid x)$ remains the same, i.e., virtual concept drift.
    \item Prior probability of class $P(\omega)$ changes while $P(x \mid \omega)$ remains the same.
\end{enumerate}
However, sometimes the class boundaries may change due to hidden contexts. This drift occurs due to 'insufficient, unknown, or observable features in a dataset' \cite{Elwell2011}. We aim to test how often such hidden contexts appear in real-time scenarios such as the automotive domain, where the environments change dynamically at a rapid pace. To this end, we employ the \gls{ph} test (see Algorithm~\ref{alg:PHtest}) to detect statistical changes in the input data stream, i.e., detect drift in data \cite{Lu2019} \cite{Baier2020}. 

\SetKw{Continue}{continue}
\SetKw{Break}{break}
\begin{algorithm}[ht]
\removelatexerror
    \KwIn{A labeled dataset $x_1$, $x_2$, $\cdots$, $x_T$}
    \algoinput{Magnitude threshold $\delta$}
    \algoinput{Detection threshold $\lambda$}
    $\Bar{x}_T \gets 1/T\sum_{t=1}^{T}x_t$\\
    $U_T \gets \sum_{t=1}^{T}(x_t-\Bar{x}_T-\delta)$\\
    $m_T \gets \min_{t=1,\cdots,T}(U_t)$\\
    ${PH}_T \gets U_T-m_T$\\
    \eIf{${PH}_T>\lambda$}{$\Theta \gets$ True}
    {$\Theta \gets$ False}
    \KwOut{Drift alarm $\Theta$}
    \caption{Page-Hinkley Test}
    \label{alg:PHtest}
\end{algorithm}

The \gls{ph} estimator needs a labeled dataset, a magnitude threshold ($\delta$), and a detection threshold ($\lambda$) as inputs. The magnitude threshold defines the degree to which noise is permitted in the dataset. For each subsequent timestep \emph{t}, a single data entry is fed to the \gls{ph} estimator and a cumulative error $U_T$ is calculated. The estimator raises an alarm if the cumulative error $U_T$ increases beyond the minimum cumulative error $m_T$ determined by detection threshold $\lambda$.

Since Algorithm~\ref{alg:PHtest} only returns a boolean value, we build $T$ subseries identical to the \gls{adf} test and run the \gls{ph} test against all of them. In that way, we can provide an example of detecting drifts in the input dataset as the vehicle is moving through different environments. For that example, an underlying \gls{ml} model has been trained on the dataset of the suburban region.  The \gls{ph} estimator's magnitude threshold $\delta$ is set to twice the standard deviation, 2$\sigma$ of the training dataset. The dataset captured by \gls{minipc} 3 from Vehicle 3 in the measurement campaign is used to test for drifts. The estimator reacts to the statistical changes in the input data and raises an alarm when a drift is detected (see Table~\ref{tab::concept_drift}). The number of detected drifts on the highway is higher by a factor of 3.3 than that of the suburban region.

\begin{table}[htb]
    \centering
    \caption{Detected drifts in test data streams for a model trained on the suburban region.}
    \label{tab::concept_drift}
    \begin{tabular}{c c c c}
    \toprule
    \multirow{2}{2cm}{\centering \bfseries Training region} & 
    \multirow{2}{2cm}{\centering \bfseries Standard deviation} & 
    \multicolumn{2}{c}{\bfseries Test region}\\ 
    \cmidrule(lr){3-4}
     && {\bfseries Highway} & {\bfseries Suburban} \\ \cmidrule(lr){1-4}
    Suburban & $2\sigma$ & 56 & 17\\
    \bottomrule
\end{tabular}
\end{table}

\begin{figure}
    \centering
    \begin{subfigure}[b]{\columnwidth}
        \centering
        \def\svgwidth{\linewidth}
        \scriptsize     
        \includegraphics[width=\linewidth]{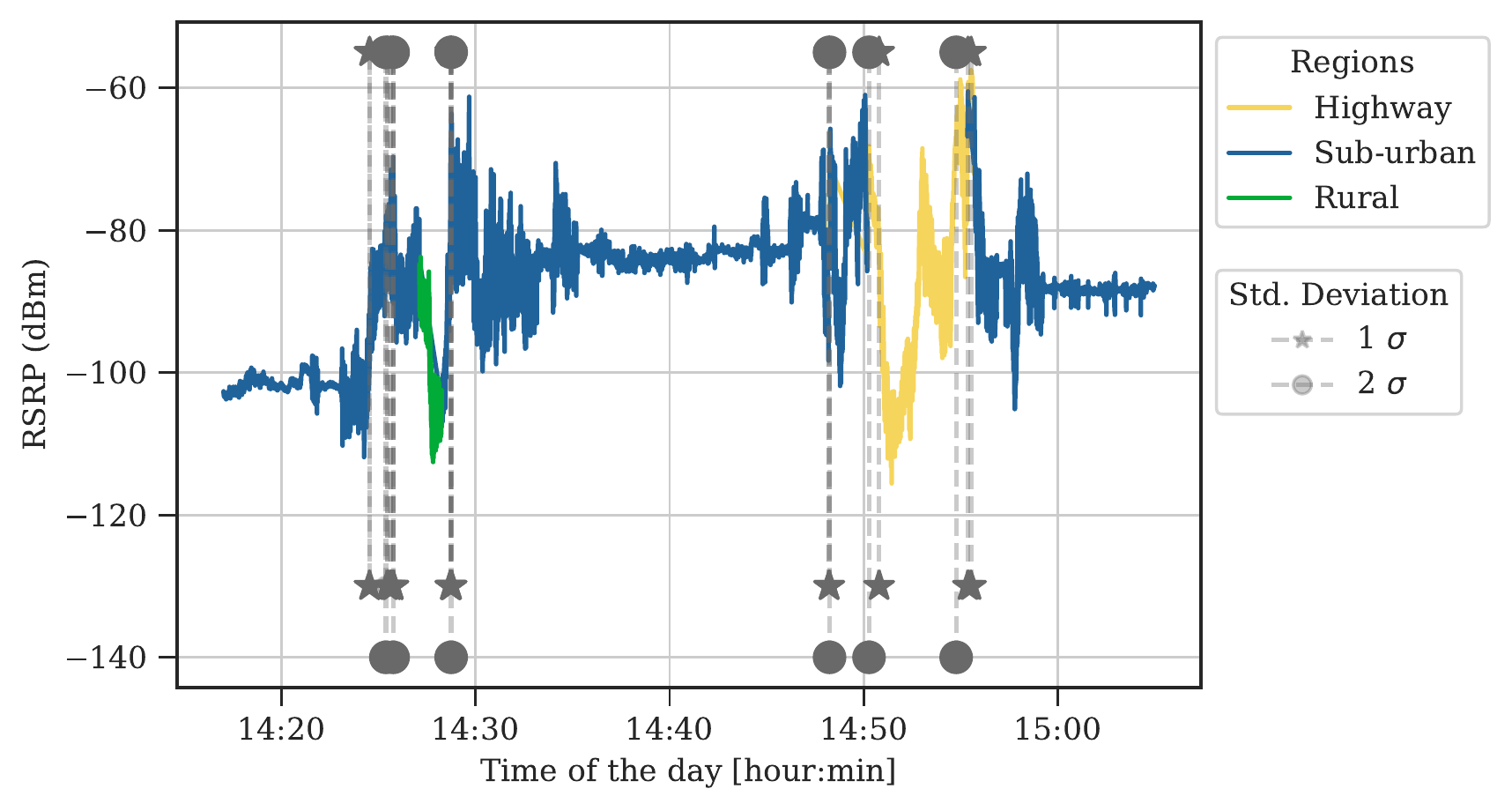}
        \caption{Suburban}
        \label{fig:drift_suburban}
    \end{subfigure}
    \hfill
    \begin{subfigure}[b]{\columnwidth}
        \centering
        \def\svgwidth{\linewidth}
        \scriptsize    
        \includegraphics[width=\linewidth]{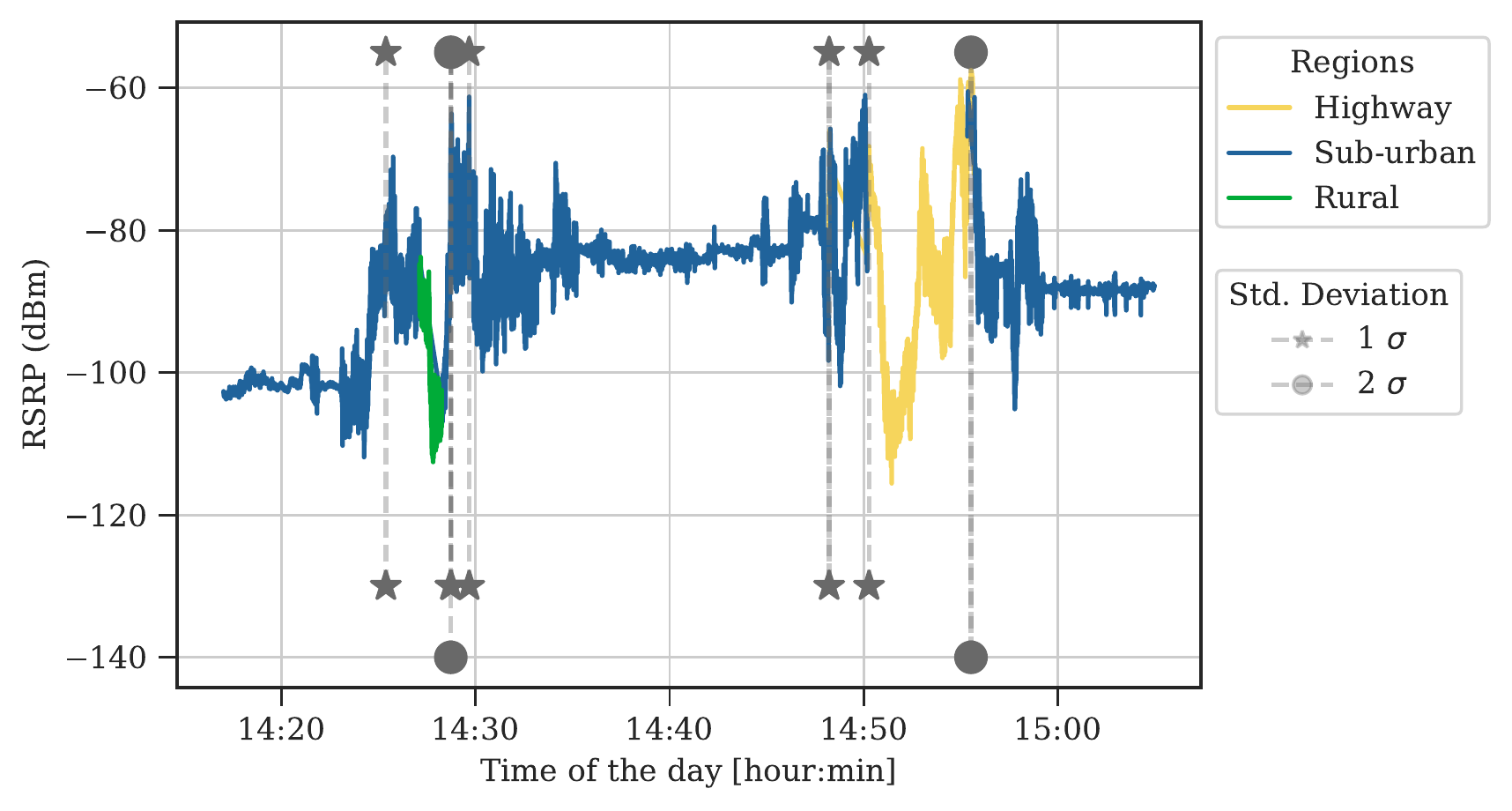}
        \caption{Highway}
        \label{fig:drift_highway}
    \end{subfigure}
    \caption{Detecting data drifts in distinct radio environments.}
    \label{fig:datadrift}
\end{figure}

To verify the time and position of these drifts, an example from the dataset, where the vehicle moves through different regions is taken (see Fig.~\ref{fig:datadrift}). Two distinct regions, i.e., suburban and highway are chosen for training the \gls{ph} estimator. Furthermore, the drifts are detected for different magnitude thresholds of $1\sigma$ and $2\sigma$ for the two training datasets. 
A change in the radio environment, when trained on the suburban dataset (see Fig.~\ref{fig:drift_suburban}), can be detected when the two different thresholds are applied.
We also got similar results when we trained for highway environments and the UE was moving between suburban, rural, and highway environments (see Fig. \ref{fig:drift_highway}).

The different types of concept drifts are expected to deteriorate \gls{ml} performance. Detecting these drifts and dealing with them as early as possible will result in a resilient and accurate \gls{ml} model. One way to handle the drifts is to include all the possible scenarios in the training dataset that a \gls{ml} model might face during its life cycle. As this might be hard, and even unrealistic in many cases, another approach is online training,
which may be based on concept drift detection~\cite{Ditzler2013,Wang2018}.
Hernangómez et. al~\cite{hernangomez2022online} provide some preliminary results on this dataset using said approach.

\subsection{Train/Test Split and Evaluation of ML Algorithms}
This section provides more details on splitting strategies, and it explains how different techniques can provide different insights. If not used properly, the splitting technique might give skewed results, and we hope to explain some of their fundamental strengths and weaknesses in the following.

\begin{figure}[tbp]
\begin{center}
\includegraphics[trim=10.1cm 4.55cm 14.4cm 5.3cm, width=\columnwidth, clip]{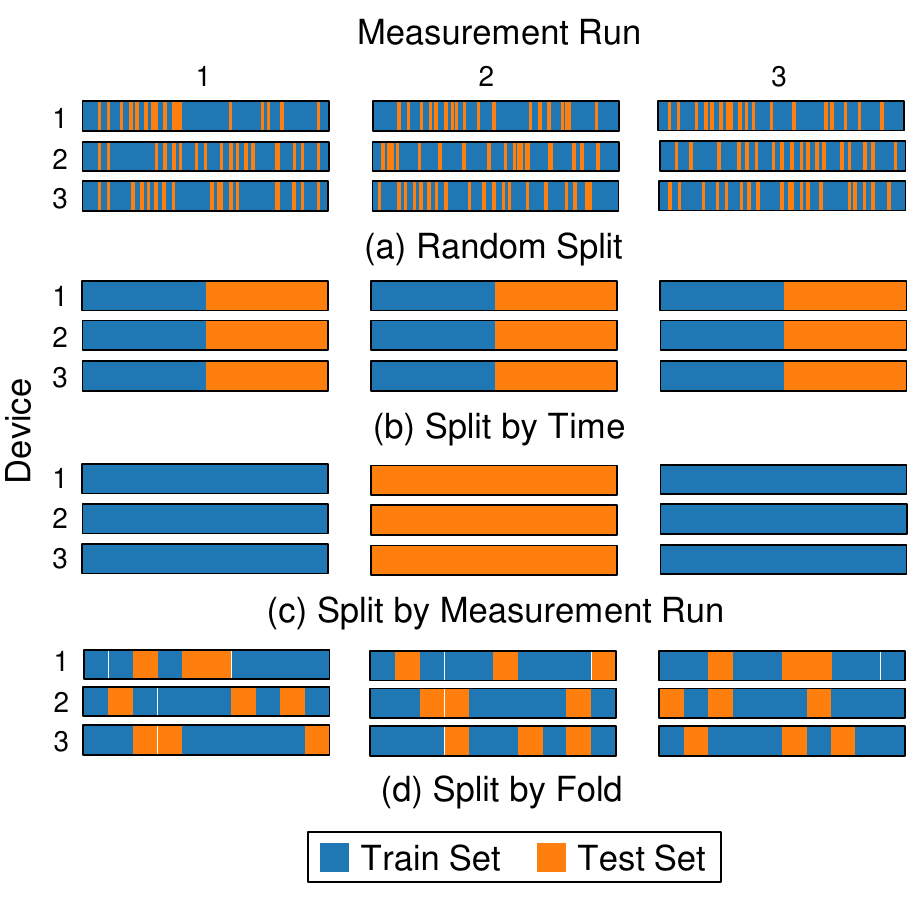}
\end{center}
\caption{Visualization of different split strategies.}
\label{fig::splits_visualization}
\end{figure}

The train/test splits included in this study are random split, split by time, split by measurement run, and split by fold. Fig.~\ref{fig::splits_visualization} provides the reader with a graphical explanation of the different splitting strategies. Subsequently, we describe the four strategies.

\paragraph*{Random split} Each dataset sample has a nonzero probability of being added to the training respectively test set. This often leads to two consecutive samples from the same device being added to the training and test set. These samples, as discussed above, share similarities, explained by their correlation structure. In the coming sections, we use a split where \SI{70}{\%} of the data are in the training dataset and \SI{30}{\%} in the testing dataset. Note that this split strategy assumes samples to be \gls{iid}, an assumption we have shown to be violated in the radio environment. 

\paragraph*{Split by time} The dataset is split into two same-size parts based on the time domain. The first part of each measurement run (independent of the device) is added to the training set, and the second to the test set. As one measurement run consisted of driving the highway segment in two directions, this represents adding one direction of travel to the training, and the second to the test set. Hence, the training and test sets contain very similar radio environments but are uncorrelated in time.

\paragraph*{Split by measurement run} A measurement run (independent of the device) is considered part of either the training or the test set. The assignment and balancing of training and test set can be challenging, as the train and test datasets need to contain similar characteristics. This splitting scheme is best used when the model's generalization performance is the main focus. It is difficult to capture all dynamics (i.e., the parameter permutations described in Section~\ref{sec::measurement_campaign}) in both training and test set. Bad performance on the test set results from the fact that the complexity and dynamics of the region are not reflected by any subset of the measurement runs. In our dataset, we ended up with a split where about \SI{70}{\%} of the data are in the training dataset, and \SI{30}{\%} in the test dataset to ensure that training contains most of the dynamics found in the test dataset.

\paragraph*{Split by folds} The last proposed splitting method aims to combine the split by time and random splitting. We divide the time domain into ten subsets of equal length for each measurement run and device. 
This approach promises a realistic prediction performance evaluation, as it negates many of the problems of random splitting, and combines advantages of the time splitting, where the models learn characteristics from the time-varying radio environment. Here, we assigned \SI{70}{\%} of the data to the training and \SI{30}{\%} to the test dataset.

\begin{figure*}[tbp]
\centering
\begin{subfigure}[b]{0.48\textwidth}
    \begin{center}
    \begin{tikzpicture}
    \begin{axis}[axis on top, grid=both,
        ymin=-4.5,ymax=7.5,
        xmin=-7.5, xmax=14.5,
    	xlabel shift = -3 pt, ylabel shift = -4 pt,
        xlabel={1\textsuperscript{st} principal component}, ylabel={2\textsuperscript{nd} principal component}, width=8cm, height=5.4cm, 
        label style={font=\scriptsize}, tick label style={font=\scriptsize},
    	legend style={
                    legend cell align=left, row sep=-0.07cm, inner ysep = 0cm, at={(0.992,0.012)}, anchor=south east,column sep=5pt}
        ] 
    	\addlegendimage{color=plot_color1,mark=*,only marks, mark size=1.75, fill opacity=0.5, draw opacity=1}
    	\addlegendimage{color=plot_color2,mark=*,only marks, fill opacity=0.5, draw opacity=1, mark size=1.75}
    	\addplot+[thick, color=blue, on layer=axis background, forget plot] graphics[xmin=-11.166,ymin=-6.5,xmax=18.166,ymax=9.5] 	{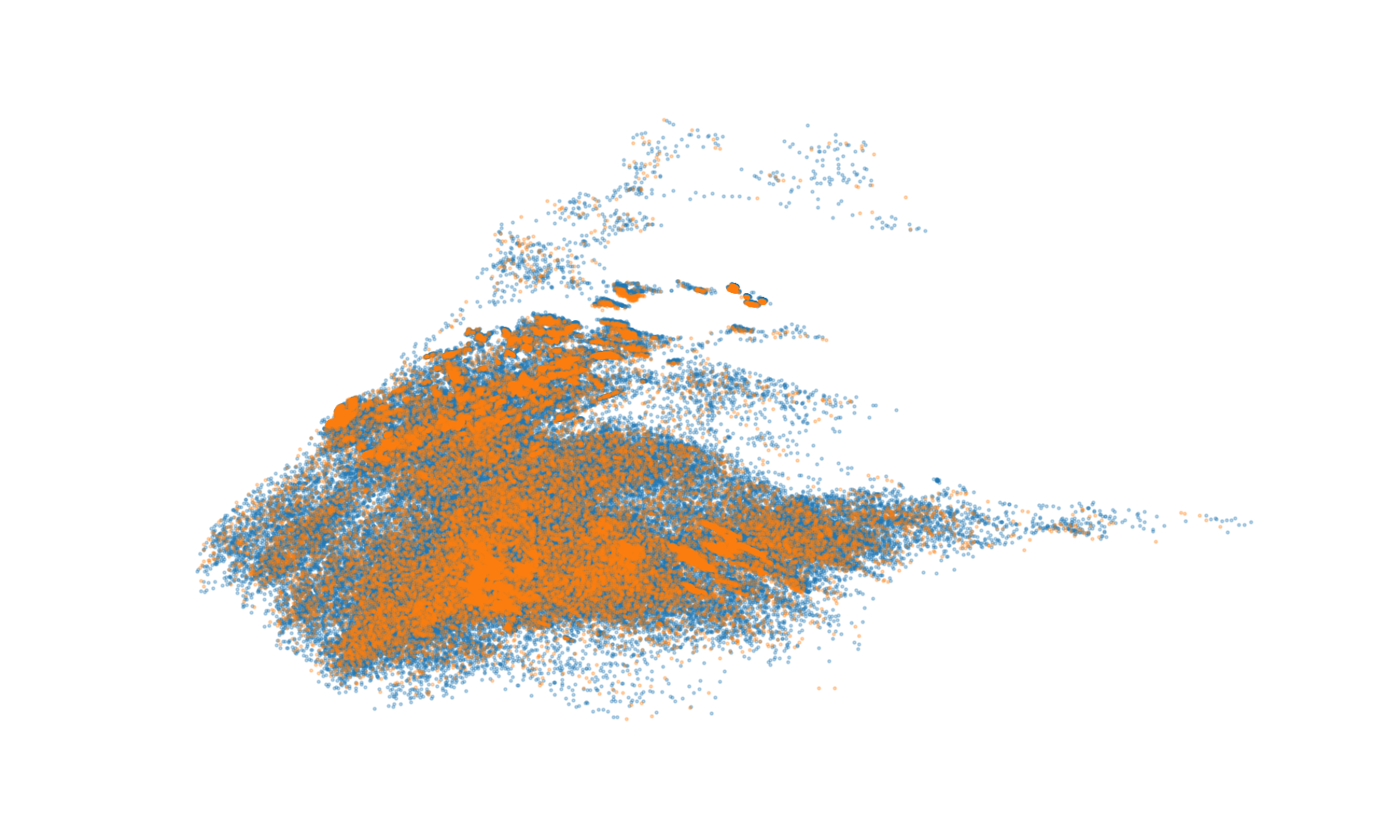}; 
    \legend{\scriptsize{Train Set}, \scriptsize{Test Set}}
    \end{axis}
    \end{tikzpicture}
    \end{center}
    \caption[]%
    {{\small Random Split}}    
    \label{fig:pca_random_split}
\end{subfigure}
\hfill
\begin{subfigure}[b]{0.48\textwidth}  
    \begin{center}
    \begin{tikzpicture}
    \begin{axis}[axis on top, grid=both,
        ymin=-4.5,ymax=7.5,
        xmin=-7.5, xmax=14.5,
    	xlabel shift = -3 pt, ylabel shift = -4 pt,
        xlabel={1\textsuperscript{st} principal component}, ylabel={2\textsuperscript{nd} principal component}, width=8cm, height=5.4cm, 
        label style={font=\scriptsize}, tick label style={font=\scriptsize},
    	legend style={
                    legend cell align=left, row sep=-0.07cm, inner ysep = 0cm, at={(0.992,0.012)}, anchor=south east,column sep=5pt}
        ] 
    	\addlegendimage{color=plot_color1,mark=*,only marks, mark size=1.75, fill opacity=0.5, draw opacity=1}
    	\addlegendimage{color=plot_color2,mark=*,only marks, fill opacity=0.5, draw opacity=1, mark size=1.75}
    	\addplot+[thick, color=blue, on layer=axis background, forget plot] graphics[xmin=-11.166,ymin=-6.5,xmax=18.166,ymax=9.5] 	{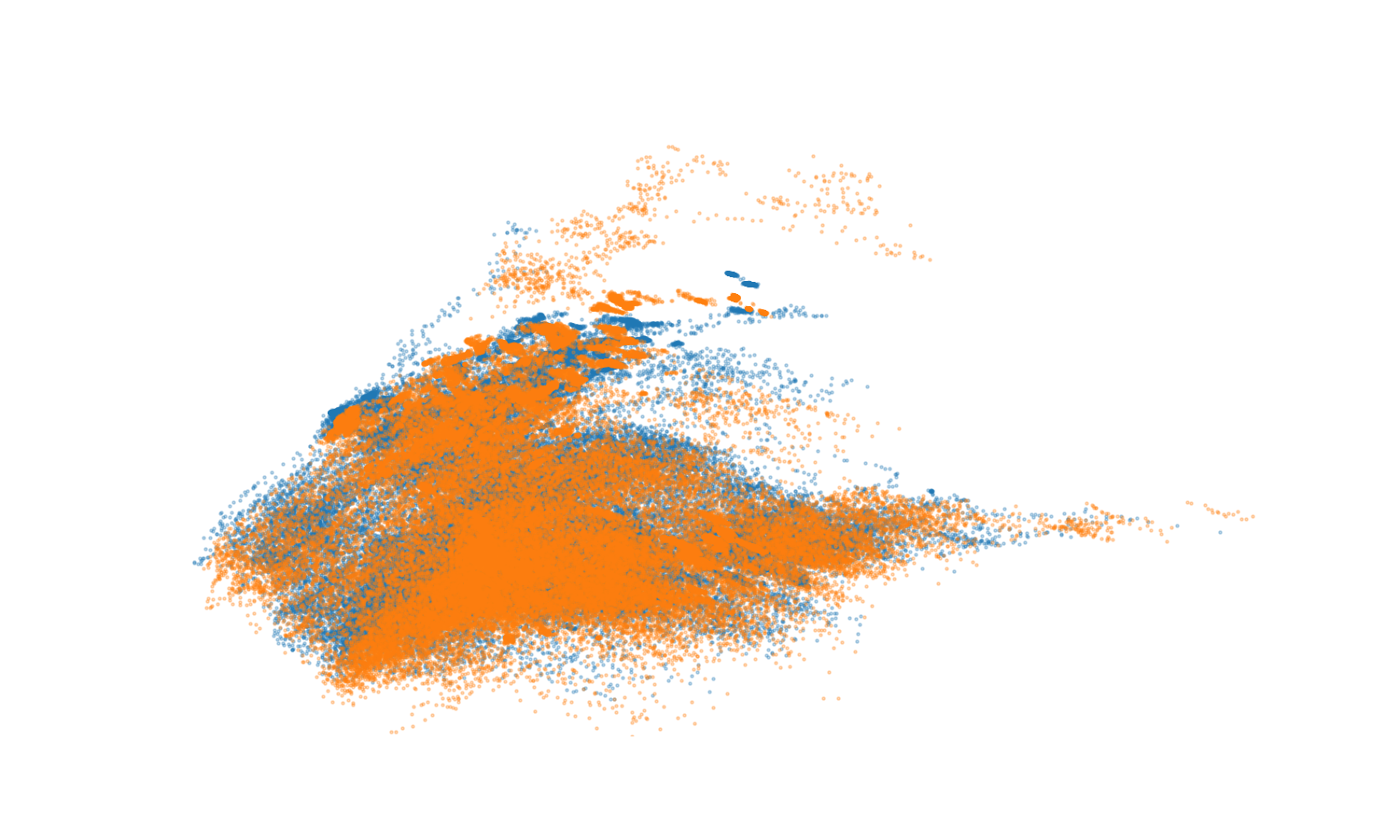}; 
    \legend{\scriptsize{Train Set}, \scriptsize{Test Set}}
    \end{axis}
    \end{tikzpicture}
    \end{center}
    \caption[]%
    {{\small Split by Time}}    
    \label{fig:pca_split_by_time}
\end{subfigure}
\vskip\baselineskip
\begin{subfigure}[b]{0.48\textwidth}   
    \begin{center}
    \begin{tikzpicture}
    \begin{axis}[axis on top, grid=both,
        ymin=-4.5,ymax=7.5,
        xmin=-7.5, xmax=14.5,
    	xlabel shift = -3 pt, ylabel shift = -4 pt,
        xlabel={1\textsuperscript{st} principal component}, ylabel={2\textsuperscript{nd} principal component}, width=8cm, height=5.4cm, 
        label style={font=\scriptsize}, tick label style={font=\scriptsize},
    	legend style={
                    legend cell align=left, row sep=-0.07cm, inner ysep = 0cm, at={(0.992,0.012)}, anchor=south east,column sep=5pt}
        ] 
    	\addlegendimage{color=plot_color1,mark=*,only marks, mark size=1.75, fill opacity=0.5, draw opacity=1}
    	\addlegendimage{color=plot_color2,mark=*,only marks, fill opacity=0.5, draw opacity=1, mark size=1.75}
    	\addplot+[thick, color=blue, on layer=axis background, forget plot] graphics[xmin=-11.166,ymin=-6.5,xmax=18.166,ymax=9.5] 	{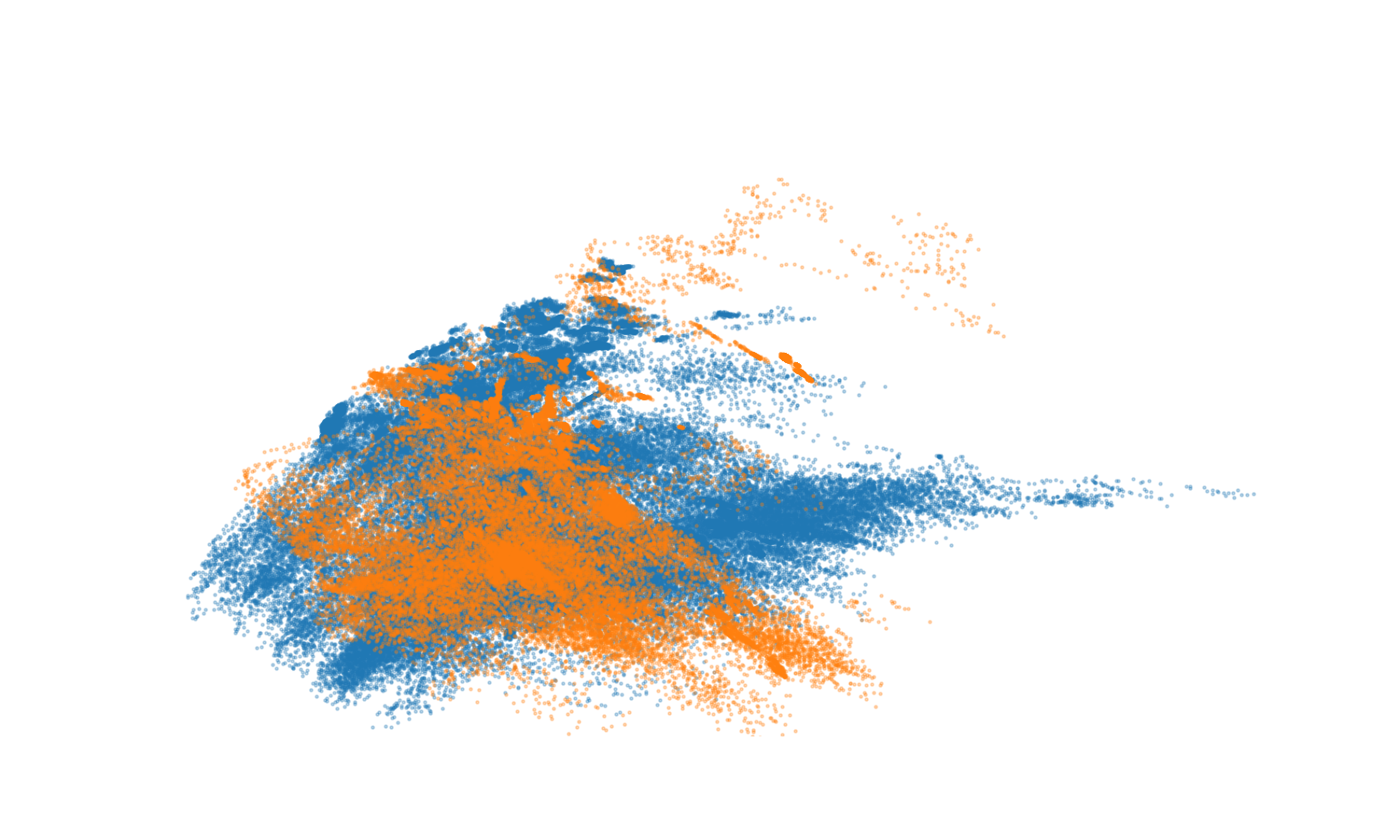}; 
    \legend{\scriptsize{Train Set}, \scriptsize{Test Set}}
    \end{axis}
    \end{tikzpicture}
    \end{center}
    \caption[]%
    {{\small Split by Measurement Run}}    
    \label{fig:pca_split_by_meas}
\end{subfigure}
\hfill
\begin{subfigure}[b]{0.48\textwidth}   
    \begin{center}
    \begin{tikzpicture}
    \begin{axis}[axis on top, grid=both,
        ymin=-4.5,ymax=7.5,
        xmin=-7.5, xmax=14.5,
    	xlabel shift = -3 pt, ylabel shift = -4 pt,
        xlabel={1\textsuperscript{st} principal component}, ylabel={2\textsuperscript{nd} principal component}, width=8cm, height=5.4cm, 
        label style={font=\scriptsize}, tick label style={font=\scriptsize},
    	legend style={
                    legend cell align=left, row sep=-0.07cm, inner ysep = 0cm, at={(0.992,0.012)}, anchor=south east,column sep=5pt}
        ] 
    	\addlegendimage{color=plot_color1,mark=*,only marks, mark size=1.75, fill opacity=0.5, draw opacity=1}
    	\addlegendimage{color=plot_color2,mark=*,only marks, fill opacity=0.5, draw opacity=1, mark size=1.75}
    	\addplot+[thick, color=blue, on layer=axis background, forget plot] graphics[xmin=-11.166,ymin=-6.5,xmax=18.166,ymax=9.5] 	{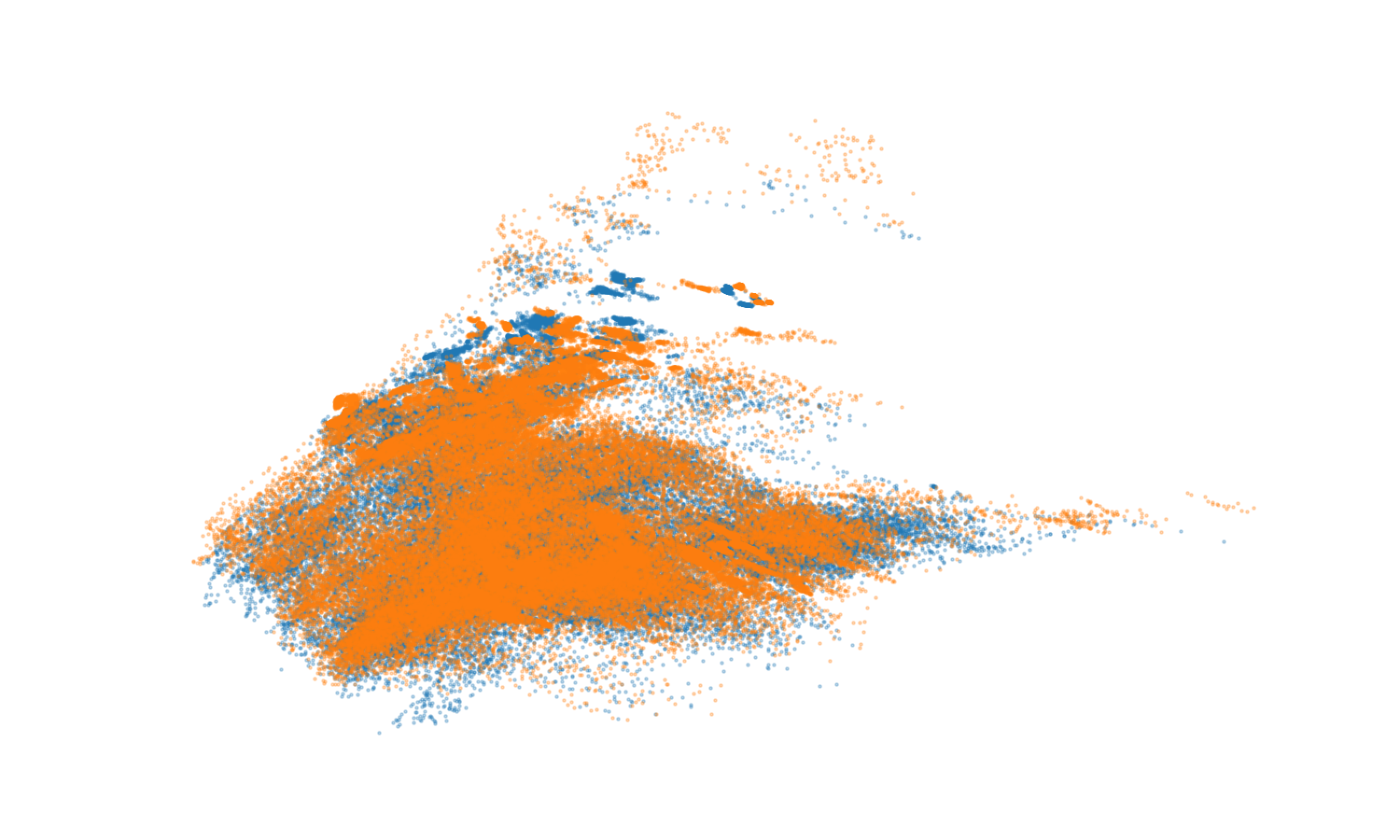}; 
    \legend{\scriptsize{Train Set}, \scriptsize{Test Set}}
    \end{axis}
    \end{tikzpicture}
    \end{center}
    \caption[]%
    {{\small Split by Fold}}    
    \label{fig:pca_split_by_fold}
\end{subfigure}
\caption{Visualization of the first two principal components for different train/test splits. The distribution of the test and train dataset varies a lot depending on the split strategy.} 
\label{fig:pca_train_test_split}
\end{figure*}[tbp]
To investigate deeper the effects of the different splitting strategies, we refer to \gls{pca}~\cite{shalev-shwartz2014understanding}, an orthogonal linear transformation transforming a dataset (e.g., the training or test set) to a new basis such that the dataset's greatest variance lies in the first principal components. Therefore, it can be considered as a dimensionality reduction method and considering only a subset of the original components, e.g., the first two for visualization purposes. We note that \gls{pca} captures only the second-order statistics and that there are alternative methods for visualization, such as {t-SNE}.
Fig.~\ref{fig:pca_train_test_split} shows the differences between the training and test sets for the four splits using \gls{pca}. The random split's train and test datasets have very similar distributions (cf. Fig.~\ref{fig:pca_random_split}). The reason is that the test set does not contain unseen propagation, data traffic, and driving scheme scenarios. Because time-series data from the radio environment seems highly correlated, the training and test datasets include many similar points. 

Therefore, although this seems to be the splitting scheme most used in the literature, there is a risk that reported performance is higher than for a deployed system. This should not discourage the use of random sampling. However, the data engineer needs to understand the inherent limitations of this splitting technique and be careful when reporting results.

The split by time alleviates to some extent the sample correlation, as consecutive samples belong to the same set (training or test). Moreover, the varying propagation environment, due to the changing direction of travel, and differences in speed, result in distributions that are less similar than for the random split (cf. Fig.~\ref{fig:pca_split_by_time}). Some regions, e.g., around the value five for the second principal component, are only present in the test set.  
The \gls{pca} can be considered a first test on how detailed collection procedures should be and highlights how time-varying components influence differences between the training and test sets. The split by time adds some limitations on the measurement duration as for balanced train and test sets more data than, e.g., for random split is required.

The split by measurement run is more appropriate to test model performance under varying and unseen parameters in the measurement procedure, the devices' behavior, and the network. In Fig.~\ref{fig:pca_split_by_meas}, it becomes clear that all these parameters have a substantial effect on the two datasets. Perhaps this is a good way to study the generalization performance of \gls{ml} models, as a training dataset will never comprise all potential parameter permutations. 
The \gls{ml} model makes predictions on samples that can be considered different, e.g., using our \gls{pca} evaluation.

We consider the split by folds the best splitting strategy for our dataset as it provides enough variation for ML models to be tested on unseen data while alleviating some of the disadvantages of the other splitting strategies. 

\subsection{Feature Acquisition and Availability Analysis}

We group different types of features based on the type and the measurement capability of a device into distinct feature groups, see Table \ref{tab::feature_table}. The table defines a name for the feature group, a corresponding abbreviation that is used throughout the rest of the paper, and the parameters (features) that the feature group inherits. The feature group \gls{phy} includes features describing the  radio environment, while \gls{chan} contains related features describing the channel. Feature group \gls{bs} contains parameters that are aggregated per cell. 

Vehicle information such as position, speed, and distance to the serving cell is included in the feature group \gls{veh}. Finally, the feature group \gls{rem} contains statistical information of \gls{phy} features and throughput in form of a radio environment map.
\begin{table*}[t]
\centering
\setlength{\tabcolsep}{4pt}
\caption{Overview of feature groups and their parameters.}
\label{tab::feature_table}
\begin{tabularx}{\textwidth}{llX}
\toprule
\textbf{Feature Group} & \textbf{Abbreviation} & \textbf{Parameters}\\
\cline{1-3}\Tstrut
\Acrlong{phy} & PHY & Mean RSRP, RSRQ, RSSI, SINR, RSRP margin, RSRQ margin, time since last handover\\
\Acrlong{chan} & CHAN & Channel matrix dominant rank index, mean CQI\\
\Acrlong{bs} & BS & Instantaneous DL/UL cell throughput (with one second resolution), number of active connected devices\\
\Acrlong{veh} & VEH & Position (latitude and longitude), speed, distance to serving cell (engineered from physical cell ID)\\
\Acrlong{rem} & REM & Radio environment map of PHY features \& instantaneous device DL/UL throughput\\
\bottomrule
\end{tabularx}
\end{table*}

Since the features are captured from several different network nodes, as previously shown in Table~\ref{tab::CapturedData}, it is not realistic to assume that all features are readily available at a specific entity such as a vehicle or one of the network nodes. Data acquisition is associated with a cost function that is not constant for different sets of features. Also, confidentiality and privacy concerns could hinder the accessibility to specific features.
This should be taken into account for \gls{ml} models for wireless networks, instead of assuming full data availability, at least without discussing their acquisition costs.

Subsequently, we further categorized the previously defined feature groups into different access scenarios. We define these according to possible scenarios in which different sets of feature groups can be collected from exemplary entities. The defined access scenarios can be found in Table \ref{tab::access_table}. The table defines a name for each access scenario, a corresponding abbreviation that is used throughout the rest of the paper, the feature groups which can be accessed, and an exemplary entity corresponding to the access scenario.

\begin{table*}[t]
\centering
\caption{Overview of network and device access scenarios and corresponding feature groups.}
\renewcommand{\arraystretch}{1.3}
\label{tab::access_table}
\begin{tabularx}{\textwidth}{llXlX}
\toprule
\textbf{Access Scenario} & \textbf{Abbreviation} & \textbf{Access to Feature Groups} & \textbf{Exemplary Entity}\\
\cline{1-4}\Tstrut
\Acrlong{md} & MD & PHY & API of operating system (e.g., Android)\\
\Acrlong{emd} & EMD & PHY, CHAN & Specialist software (e.g., MobileInsight~\cite{li2016mobileinsight})\\
\Acrlong{mdnet} & MDNET & PHY, BS & Network operator\\
\Acrlong{mdrem} & MDREM & PHY, REM & Mapping and navigation service\\
\Acrlong{emdnet} & EMDNET & PHY, CHAN, BS & Network operator\\
\Acrlong{remnet} & REMNET & PHY, BS, REM & Network operator\\
\Acrlong{devnet} & DEVNET & PHY, BS, VEH & Car manufacturer (with MNO cooperation)\\
\Acrlong{dev} & DEV & PHY, CHAN, VEH & Car manufacturer\\
\Acrlong{full} & FULL & PHY, CHAN, BS, VEH, REM & -\\
\bottomrule
\end{tabularx}
\end{table*}
The first access scenario is \gls{md} and refers to features available at the modem, such as parameters from feature group \gls{phy}, which can often be easily accessed, e.g., via an API. \gls{emd} refers to access to more modem data provided, e.g., by specialized software.
Both \gls{md} and \gls{emd} correspond to data available at a \gls{ue}, while \gls{mdnet}, \gls{emdnet}, and \gls{remnet} are available for a network operator with varying acquisition costs (in terms of computing and signaling). For example, \gls{phy} features are reported by the \gls{ue} to the network during measurement reports, but these are usually sent after specific events are triggered. On the other hand, features from the \gls{chan} group are reported regularly if the \gls{ue} receives \gls{dll} traffic. Since the dataset is based on vehicular measurements, we also introduce access scenarios \gls{devnet} and \gls{dev} which include \gls{veh} features in addition to other feature groups. For comparison, we also include a \gls{full} access scenario, which has access to all feature groups.

The aim of this paper is not to define specific acquisition costs associated with the different access scenarios, but to acknowledge the potential trade-off between data acquisition costs and \gls{ml} performance and compare the performance in the next section.
Additionally, we would like to highlight two aspects in particular: First, there are data acquisition costs, which vary depending on the different feature groups. Second, \gls{ml} models need to be tested under different assumptions on data availability including concerns of data governance and ownership.

\subsection{ML Models for QoS Prediction}
We  briefly introduce different \gls{ml} models, that are relevant to the prediction task. These include linear regression, ensemble methods, and neural networks. In terms of completeness, we also add a statistical model that serves as a baseline.
The model hyperparameters were tuned using~\cite{optuna} with 5-fold cross-validation.
The split by folds was used to split the dataset into a train and validation set.
We used the \gls{mae} as the cost function.

\paragraph*{\Gls{rem} of the Throughput} After building a \gls{rem} of the \gls{dll}/\gls{ul} throughput, the predicted value is the interpolated throughput value for the current \gls{ue} position. This method serves as a statistical baseline.

\paragraph*{\Gls{lr}} Linear regression is a relatively simple statistical model where a dependent variable is explained by multiple independent variables multiplied by coefficients defining their weights. In a \gls{ml} task, the weights of these coefficients are learned.

\paragraph*{\Gls{rf}} A random forest is an ensemble method based on decision trees.
The trees are built independently and in a regression task, the mean of the individual trees is the prediction result.
Our \gls{rf} is composed of 793 parallel trees with a maximum depth of 19.

\paragraph*{\Gls{gb}} Gradient boosting is also an ensemble method based on decision trees.
Here, the trees are built sequentially, minimizing the loss function with each new tree.
The weighted mean according to the trees' prediction performance is the prediction result.
Our \gls{gb} ensemble is composed of 715 sequential trees with a maximum depth of 10.

\paragraph*{\Gls{mlp}} 
An \gls{mlp} is a feedforward neural network that stacks together multiple layers of neurons in a sequential architecture.
It is a \gls{dl} model that has been successfully applied to regression problems with labeled tabular datasets.
Our \gls{mlp} consists of four hidden layers 256, 128, 64, and 32 neurons each and rectified linear unit (ReLU) activation functions.
It is trained with an Adam optimizer, with a learning rate of 0.001, and a batch size of 16.
The parameters of the neural network are initialized randomly.
For regularization, we use early stopping with the criterion that training ends when the error does not decrease for eight epochs on a validation dataset.
\section{Throughput Prediction}\label{sec::ThroughputPrediction}
In this section, we discuss the throughput prediction task based on \gls{ml}. The prediction task we picked is to estimate the achieved maximum throughput under a high network load. The high network load conditions mean that the end users have to share the network resources.

We start by investigating the performance of different \gls{ml} models and continue by looking at the performance when different data access scenarios apply. We then compare the influence of the split strategies on the training and test datasets. We continue looking at the effects of sampling on the prediction task. Subsequently, we give an example of how much concept drift can deteriorate performance. We conclude by showing that \gls{gnet} can play an important role in \gls{ml}-related tasks.

For all the provided results we report multiple metrics~\cite{metrics_applied}, as this allows a more flexible evaluation of the prediction performance for diverse use cases and when applicable we discuss the limitations of specific metrics. 

\subsection{Model Performance}

In Tables~\ref{tab::reg_metrics_dl} and~\ref{tab::reg_metrics_ul}, we compare different models on standard regression metrics for the \gls{dll} and \gls{ul} direction, respectively. We assume to use all the information available for the prediction (i.e., \gls{full}), consider all environments jointly, and refer to split by folds.
Moreover, we relied on all available data samples for performance evaluation.

In each table, we highlight the best-performing model per metric in bold. For the \gls{dll} direction (Table~\ref{tab::reg_metrics_dl}) the \gls{gb} outperforms the other models for all metrics. \gls{rf} consistently shows slightly lower performance than \gls{gb}. For the \gls{mape}, the \gls{mlp} is on par with \gls{gb}, but it shows worse performance than \gls{gb} and \gls{rf} for most other metrics. Considering the two simpler models, \gls{lr} outperforms the \gls{rem} interpolation for most metrics, underlining that the use of context information, which is included in the input features, is crucial for the prediction task. Moreover, as the feature space is not linear, the \gls{lr} performance is poor. 

The \gls{ul} direction (Table~\ref{tab::reg_metrics_ul}) mainly corresponds to the performance observations in the \gls{dll} prediction task. We observe that the absolute gap between \gls{gb} and \gls{rf} narrows for \gls{mae}, \gls{medae}, and \gls{rmse} (due to lower \gls{ul} data rates), with \gls{rf} showing similar performance as \gls{gb} for the R\textsuperscript{2} score. The \gls{ul} throughput range (up to 23 Mbps) is smaller than the \gls{dll} throughput range (up to 70 Mbps), which is reflected in the prediction performance. While the average \gls{mae} in UL is significantly smaller than the average \gls{mae} in \gls{dll}, the \gls{mape} of the \gls{dll} is comparably lower than in \gls{ul}.

As there is a trade-off between maximum performance and acquisition cost, we found that performance metrics, such as \gls{mae}, converges at around 5000 training samples. Adding further samples only leads to minor increases in prediction accuracy. However, we note that this result is specific to our dataset and may not generalize to other datasets because the minimum required dataset size depends on a multitude of factors. Such factors are the environment dynamics, the network dynamics, and the specific use case (i.e. required accuracy). Nevertheless, this result shows that the performance return degrades with an increasing number of samples and measurement campaigns should be planned carefully as there is space for optimizing the data acquisition expenditures.

In summary, \gls{gb} proved to be the best-performing model considering both data traffic directions and all metrics. At the same time, the interested reader should note that there are a plethora of \gls{ml} algorithms that can be applied and provide good performance. The \gls{ml} engineer can always pick one based on criteria like the model's complexity and the degrees of explainability. As the ensemble methods based on decision trees offer a good degree of explainability and robustness against outliers, we pick \gls{gb} for the following sections.

\begin{table}[htb]
    \centering
    \caption{Performance comparison of different regression models in \gls{dll} direction.}
    \label{tab::reg_metrics_dl}
    \begin{tabular}{lrrrrr}
\toprule
{\bfseries Model} & {\bfseries MAE} & {\bfseries MAPE} & {\bfseries MedAE} & {\bfseries RMSE} & \textbf{R\textsuperscript{2}} \\
\midrule
RF & 2.61 & 0.31 & 1.67 & 3.91 & 0.92 \\
GB & \bfseries 2.46 & \bfseries 0.29 & \bfseries 1.48 & \bfseries 3.77 & \bfseries 0.93 \\
LR & 4.91 & 0.78 & 3.57 & 6.70 & 0.77 \\
MLP & 2.71 & \bfseries 0.29 & 1.59 & 4.23 & 0.91 \\
REM(y) & 6.80 & 0.84 & 4.26 & 9.97 & 0.53 \\
\bottomrule
\end{tabular}

\end{table}
\begin{table}[htb]
    \centering
    \caption{Performance comparison of different regression models in \gls{ul} direction.}
    \label{tab::reg_metrics_ul}
    \begin{tabular}{lrrrrr}
\toprule
{\bfseries Model} & {\bfseries MAE} & {\bfseries MAPE} & {\bfseries MedAE} & {\bfseries RMSE} & \textbf{R\textsuperscript{2}} \\
\midrule
RF & 1.13 & 0.50 & 0.79 & 1.63 & \bfseries 0.90 \\
GB & \bfseries 1.08 & \bfseries 0.46 & \bfseries 0.73 & \bfseries 1.58 & \bfseries 0.90 \\
LR & 2.31 & 1.79 & 1.88 & 2.96 & 0.66 \\
MLP & 1.20 & 0.53 & 0.75 & 1.80 & 0.87 \\
REM(y) & 3.20 & 0.96 & 2.41 & 4.28 & 0.25 \\
\bottomrule
\end{tabular}

\end{table}

\subsection{Access Scenarios \& Feature Groups}
We continue by looking at a different set of input features and how much these affect the achieved performance. In the literature, input features are very often considered quite alike in terms of acquisition costs. In a real network, there is a cost associated with collecting data from different nodes and end-users.
We compare the different prediction performances for the previously defined access scenarios, from Section~\ref{sec::MLforRadio}, in Tables \ref{tab::feature_groups_DL} and \ref{tab::feature_groups_UL} for \gls{dll} and \gls{ul} direction, respectively.

In both \gls{dll} and \gls{ul}, the performance of different access scenarios is very similar, with the only difference that, in general, for \gls{dll} the \gls{mape} is significantly lower compared to \gls{ul}, and the \gls{mae}, \gls{medae}, and \gls{rmse} are higher for \gls{dll} due to the larger value range.
The worst-performing access scenario in both cases is \gls{md}, where only the feature group \gls{phy} is used. 
Prediction for \gls{emd}, \gls{mdrem}, and \gls{dev} performs slightly better but not by a large margin. This probably indicates that additional channel or statistical/historical information might be beneficial but does not add much useful information in addition to physical layer parameters, which can be obtained easily. The prediction with only \gls{ue} features (\gls{md}/\gls{emd}) achieves the worst performance across all presented metrics and access scenarios.
However, adding  \gls{bs}  features, demonstrated in access scenarios \gls{mdnet}, \gls{emdnet}, \gls{remnet} and \gls{devnet}, significantly improves the prediction performance, resulting in a halved prediction error for \gls{dll}. The same holds true for \gls{ul} direction except for the \gls{mape}, which is still significantly lower by around \si{20-30}\, percentage points. Also, the R\textsuperscript{2} score for both \gls{dll} and \gls{ul} is increased to approximately 0.9 once \gls{bs} features are added.

The best prediction performance is achieved when access to all features is provided, which is emphasized by the bold rows for access scenario \gls{full}. As mentioned in the previous section, we introduced this for comparison only, as access to all features is often not realistic or associated with high acquisition costs. However, we see that the performance of \gls{full} is improved by only a small margin compared to the access scenarios including feature group \gls{bs} - even compared to \gls{mdnet}, which contains only simple feature group \gls{phy} in addition to feature group \gls{bs}. Thus, \gls{mdnet} might be a feasible candidate if one tries to find a good trade-off between better performance and lower data acquisition costs. 
\begin{table}[]
\centering
\caption{Performance  comparison of different access scenarios with access to different feature groups in DL using the split by folds.}
\label{tab::feature_groups_DL}
\setlength{\tabcolsep}{4pt}
\begin{tabular}{lrrrrr}
\toprule
{\bfseries Access scenario} & {\bfseries MAE} & {\bfseries MAPE} & {\bfseries MedAE} & {\bfseries RMSE} & \textbf{R\textsuperscript{2}} \\
\midrule
MD & 6.21 & 0.62 & 3.86 & 9.27 & 0.56 \\
EMD & 5.54 & 0.56 & 3.43 & 8.27 & 0.66 \\
MDNET & 2.93 &  0.32 &  1.86 &  4.38 &   0.91 \\
MDREM & 5.81 & 0.53 & 3.46 & 8.85 & 0.62 \\
EMDNET & 2.56 & 0.30 & 1.57 & 3.91 & 0.92 \\
REMNET & 2.78 & 0.30 & 1.69 & 4.21 & 0.91 \\
DEVNET & 2.68 & \bfseries 0.29 & 1.65 & 4.07 & 0.92 \\
DEV & 5.01 & 0.49 & 2.79 & 7.96 & 0.68 \\
FULL & \bfseries 2.46 & \bfseries 0.29 & \bfseries 1.48 & \bfseries 3.77 & \bfseries 0.93 \\
\bottomrule
\end{tabular}

\end{table}

\begin{table}[]
\centering
\caption{Performance  comparison of different access scenarios with access to different feature groups in UL using the split by folds.}
\label{tab::feature_groups_UL}
\setlength{\tabcolsep}{4pt}
\begin{tabular}{lrrrrr}
\toprule
{\bfseries Access scenario} & {\bfseries MAE} & {\bfseries MAPE} & {\bfseries MedAE} & {\bfseries RMSE} & \textbf{R\textsuperscript{2}} \\
\midrule
MD & 2.93 & 0.97 & 2.12 & 4.03 & 0.38 \\
EMD & 2.68 & 0.89 & 1.90 & 3.72 & 0.45 \\
MDNET &  1.18 &  0.65 &  0.80 &  1.74 &  0.88 \\
MDREM & 2.62 & 0.73 & 1.89 & 3.63 & 0.48 \\
EMDNET & 1.15 & 0.72 & 0.77 & 1.70 & 0.89 \\
REMNET & 1.12 & 0.50 & 0.75 & 1.64 & \bfseries 0.90 \\
DEVNET & 1.14 & 0.56 & 0.80 & 1.66 & 0.89 \\
DEV & 2.48 & 0.76 & 1.75 & 3.49 & 0.53 \\
FULL & \bfseries 1.08 & \bfseries 0.46 & \bfseries 0.73 & \bfseries 1.58 & \bfseries 0.90 \\
\bottomrule
\end{tabular}

\end{table}

\subsection{Split Strategies}
As discussed in Section~\ref{sec::MLforRadio} there are different approaches for splitting the dataset into train and test sets.

We present results for using different splitting strategies for the DL direction in Table~\ref{tab::reg_by_split_dl}.
According to all analyzed metrics, the random split, as described above, yields the best prediction results. 

The performance would likely decrease drastically during deployment, compared to other splits, when the model is presented with new data with a different correlation structure. 

In our case, the performance of the random split is \SI{56}{\%} better compared to the split by folds.
However, the performance of the random split is likely overly optimistic due to the temporal correlations  between samples~\cite{cv_correlated}.
Hence, ML models trained with the random split strategy might perform worse on new uncorrelated data that appears in real deployments.

As reasoned earlier, performance results obtained with the split by folds are more likely to be more robust to other deployment scenarios.
The split by folds achieves the second-best results for most metrics.

On the other hand, the performance of the splits by time and split by measurement perform worse, which supports the hypothesis that there is rather a change in the statistics over time. This change can be attributed to the radio environment, as there are different numbers and types of vehicles around during the data collection process.

The results above highlight the need for picking a single or a multitude of splitting results for assessing the \gls{ml} performance for a specific use case. Failure to have a clear splitting strategy might lead to a false estimation of the performance during the deployment phase.

\begin{table}[htb]
    \centering
    \caption{Performance of GB for different splits for \gls{dll}.}

    \label{tab::reg_by_split_dl}
    \begin{tabular}{lrrrrr}
\toprule
{\bfseries Split} & {\bfseries MAE} & {\bfseries MAPE} & {\bfseries MedAE} & {\bfseries RMSE} & \textbf{R\textsuperscript{2}} \\
\midrule
random & \bfseries 1.58 & \bfseries 0.18 & \bfseries 0.88 & \bfseries 2.62 & \bfseries 0.97 \\
time & 3.66 & 0.57 & 2.70 & 5.10 & 0.86 \\
meas & 4.50 & 0.24 & 3.44 & 6.03 & 0.88 \\
folds & 2.46 & 0.29 & 1.48 & 3.77 & 0.93 \\
\bottomrule
\end{tabular}

\end{table}

\begin{figure}
\begin{center}
\includegraphics[width=\linewidth]{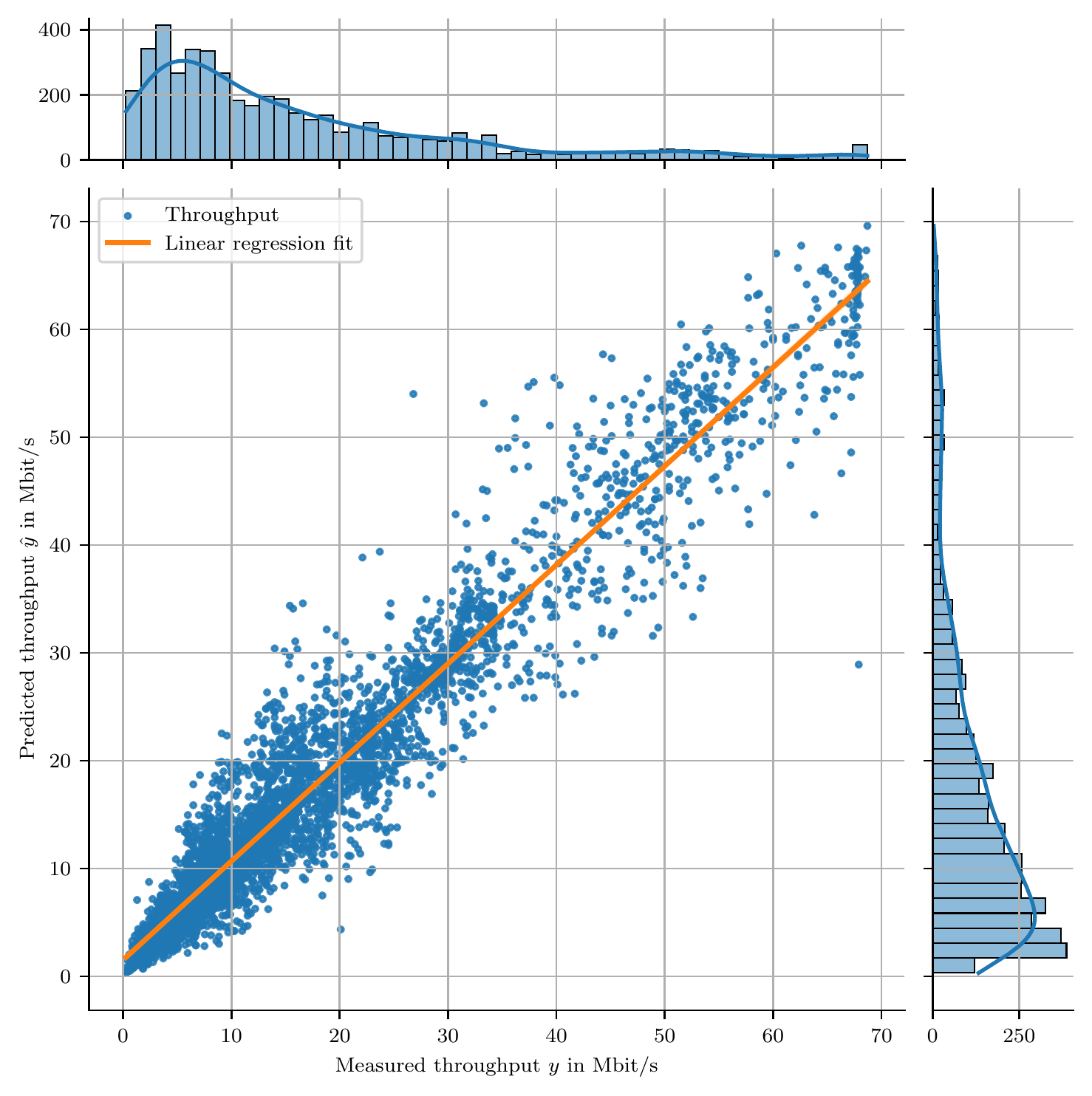}
\end{center}
\caption{Visualization of the predicted throughput over the given throughput.}
\label{fig::y_pred_y_test}
\end{figure}

In Fig.~\ref{fig::y_pred_y_test}, we show the predicted values against the measured ones. The \gls{dll} throughput was predicted with the GB model, utilizing the features contained in the \gls{full} access type and using the split by folds. 

In the figure, we also include two histograms for both axes. As multiple devices were competing for higher throughput, the total capacity of the network was typically split between multiple users making the highest throughput measurements a rarity. That well explains the skewed nature of the distribution. The interested reader should note that the mean type of metrics used in this section, \gls{mae} and \gls{mape}, are heavily influenced by the long distribution tails. Therefore we have introduced median-based metrics, such as the \gls{medae}, that are less influenced by a few large outliers.

\subsection{Sampling Intervals}

As discussed in Sections~\ref{sec::StatisticalProperties} and ~\ref{sec::MLforRadio}, the radio environment has a correlation structure that renders consecutive measurements quite alike. That means that slower sampling intervals might be used to reduce overhead, such as battery consumption and signaling. To showcase that, we compare in Table~\ref{tab::reg_downsampled} results for slower sampling intervals than the \gls{gb} results in Table~\ref{tab::reg_metrics_dl}. We see that a 50 percent slower sampling interval (\SI{2}{s}) brings a negligible drop in the  performance of an \gls{ml} model. Even sampling procedures that are 75 percent slower than the initial sampling (\SI{4}{s}) might be used and still, performance could be more than adequate for some use cases, with the biggest change being that the \gls{mae} increases by \SI{20}{\%}. We also noticed that the reduction in the sampling intervals made \gls{phy} features less relevant and increased the importance of the \gls{rem} features. 
The correlation structure of the radio environment can be further exploited to optimize sampling procedures while keeping the \gls{ml} model's performance within the requirements of a use case. A deeper understanding of the correlation structure of the radio environment can drastically benefit the sampling procedures~\cite{dyspanLondon} supporting more efficient \gls{ml} workflows.
\begin{table}[htb]
    \centering
    \caption{Performance of GB for different downsampling periods for \gls{dl}.}
    \label{tab::reg_downsampled}
    \begin{tabular}{lrrrrr}
\toprule
{\bfseries Period} & {\bfseries MAE} & {\bfseries MAPE} & {\bfseries MedAE} & {\bfseries RMSE} & \textbf{R\textsuperscript{2}} \\
\midrule
\textcolor{gray}{1s} &\textcolor{gray}{2.46} &  \textcolor{gray}{0.29} & \textcolor{gray}{1.48} & \textcolor{gray}{3.77} & \textcolor{gray} {0.93} \\
2s & \bfseries 2.66 & \bfseries 0.30 & \bfseries 1.69 & \bfseries 4.08 & \bfseries 0.91 \\
3s & 2.93 & 0.42 & 1.83 & 4.48 & 0.89 \\
4s & 3.11 & 0.37 & 1.97 & 4.68 & 0.88 \\
5s & 3.58 & 0.40 & 2.02 & 5.71 & 0.84 \\
10s & 4.40 & 0.41 & 2.77 & 6.63 & 0.82 \\
\bottomrule
\end{tabular}

\end{table}

\subsection{Prediction Horizon}
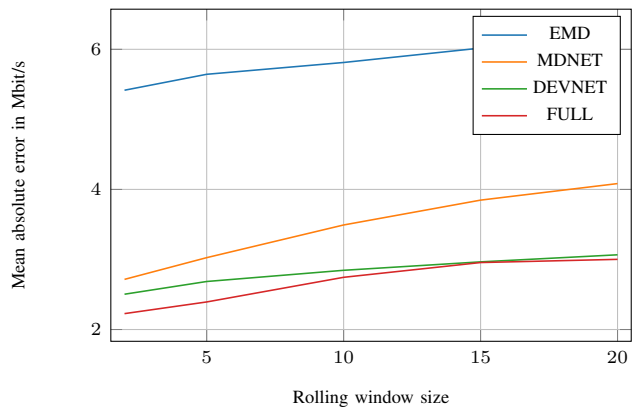
\begin{figure}[tbp]
\begin{center}
\begin{tikzpicture}
\begin{axis}[
	xmin = 1.5, xmax = 20.5,
    ylabel = Mean absolute error in Mbit/s,
    xlabel = Rolling window size,
    width=\wFig, height=\hFig, grid=both,
	xlabel shift = -3 pt, ylabel shift = -4 pt, label style={font=\scriptsize}, tick label style={font=\scriptsize}
    ]
    \addplot+[plot_color1,mark=None,semithick]%
    table [x=window, y=mean_absolute_error, col sep=comma] {data/rolling_EMD_DL.csv};
    \addplot+[plot_color2,mark=None,semithick]%
    table [x=window, y=mean_absolute_error, col sep=comma] {data/rolling_MDNET_DL.csv};
    \addplot+[plot_color3,mark=None,semithick]%
    table [x=window, y=mean_absolute_error, col sep=comma] {data/rolling_DEVNET_DL.csv};
    \addplot+[plot_color4,mark=None,semithick]%
    table [x=window, y=mean_absolute_error, col sep=comma] {data/rolling_FULL_DL.csv};
    \legend{\scriptsize{EMD}, \scriptsize{MDNET}, \scriptsize{DEVNET}, \scriptsize{FULL}}
    \end{axis}
\end{tikzpicture}
\end{center}
\vspace*{-3mm}%
\caption{Rolling window response variable.}
\label{fig::rolling_win_y}
\end{figure}

We continue by changing the prediction problem slightly. Instead of predicting the achievable instantaneous throughput, we predict the throughput in sevaral seconds defined by the prediction horizon. This is shown in Fig.~\ref{fig::rolling_win_y}, for different access scenarios. We see that the access scenario \gls{emd} provides poor performance, similar to the previous subsections. The biggest improvement in the prediction error occurs when we include the \gls{mdnet} access group, which includes the network information. The access scenario performance \gls{devnet} provides marginal improvements, with the prediction performance dropping slower for longer prediction horizons. Best performance is achieved for access scenario \gls{full}, although after a prediction horizon of 12 seconds, there is no real gain, of any extra-added features, as it performs as well as the \gls{devnet} access scenario.

\subsection{Concept Drift and ML}
As was discussed in Section~\ref{sec::StatisticalProperties}, there are concept shifts occurring frequently in the radio environment.  Here, we provide some examples of the degradation of the performance for cases that \gls{ml} faces concept drifts. Our example is shown in Table~\ref{tab::reg_inter_env}, where we trained a model on the suburban environment and afterward assumed that it was deployed on a vehicle traveling across a highway environment. We have seen already in Section~\ref{sec::StatisticalProperties} that while driving from a suburban environment onto a highway, a large number of concept drifts is detected. The performance of the \gls{ml} algorithm drops drastically, to the level of a statistical baseline as shown from the R\textsuperscript{2}.

\begin{table}[htb]
    \centering
    \caption{Performance of GB model trained on suburban environment and tested on highway environment.}
    \label{tab::reg_inter_env}
    \begin{tabular}{lrrrrr}
\toprule
{\bfseries Direction} & {\bfseries MAE} & {\bfseries MAPE} & {\bfseries MedAE} & {\bfseries RMSE} & \textbf{R\textsuperscript{2}} \\
\midrule
DL & 9.68 & 0.47 & 5.02 & 15.18 & 0.05 \\
UL & 3.50 & 0.56 & 2.33 & 5.00 & 0.03 \\
\bottomrule
\end{tabular}

\end{table}

\subsection{Consumer Grade Devices}
The presented results were calculated using data captured by \gls{minipc}s. As \gls{minipc} is relatively expensive and comes with increased processing and measurement set-up effort, we compare it in this section to data from \gls{gnet}. Since \gls{gnet} is significantly more accessible in price and set-up, the question arises to which degree the dataset differs from the \gls{minipc} dataset, and, more specifically, whether a \gls{gnet} dataset can be sufficient for the use case of throughput prediction.

For this comparison, we use data only from the vehicles that were equipped both with a \gls{gnet} and a \gls{minipc} (Vehicles 3 and 4) since this creates datasets of roughly the same size and diversity. Moreover, only feature groups \gls{phy} and  \gls{bs} were considered, (i.e., access scenario \gls{mdnet}), to match the more limited feature set availability of the \gls{gnet}. 
Table \ref{tab::gn_pc} presents a comparison of prediction results between the \gls{gnet} and the corresponding \gls{minipc}.
Both in \gls{ul} and \gls{dll}, the different metrics display similar performance. 

The R\textsuperscript{2} scores of the \gls{gnet} is slightly better than the \gls{minipc}'s. The \gls{minipc} has higher sensitivity and we have seen larger number of outliers at the higher throughput ranges. Such outliers negatively influence the  \gls{minipc}'s performance.
\begin{table}[]
\centering
\caption{Performance comparison of different devices in \gls{ul} and \gls{dl} with MDNET features.}
\label{tab::gn_pc}
\renewcommand{\arraystretch}{1.2}
\setlength{\tabcolsep}{5pt}
\begin{tabularx}{\columnwidth}{@{}llllll@{}}
\toprule
\textbf{Scenario} & \textbf{MAE} & \textbf{MAPE} & \textbf{MedAE} & \textbf{RMSE} & \textbf{R\textsuperscript{2}} \\ \midrule
UL CE (Veh. 3\&4)  & 1.13 & \bfseries 0.38 &  0.74 &  1.65 &  0.82 \\
UL DME (Veh. 3\&4) & \bfseries 0.93 &  0.41 & \bfseries 0.64 & \bfseries 1.29 &\bfseries 0.81 \\ \hline
DL CE (Veh. 3\&4)  & 3.37 &   0.34 &  2.06 &  5.17 &  0.84 \\
DL DME (Veh. 3\&4)  & \bfseries 2.57 & \bfseries 0.31 & \bfseries 1.71 & \bfseries  3.69 & \bfseries 0.82 \\
\bottomrule
\end{tabularx}
\end{table}
Overall, our findings show that \gls{gnet} can be part of a data collection procedure. More expensive devices might bring some small benefits but with diminishing returns.

\section{Interpretability and Explainability of the Models}\label{sec::interpretability}
In this section, we use \gls{SHAP}~\cite{NIPS2017_7062} and \gls{ALE} \cite{ale} for explaining how the input features affect the prediction of the \gls{ml} model.  

\subsection{Shapley Additive Explanations}
\gls{SHAP} is a framework to explain individual predictions. We use the model-agnostic variant of SHAP as described in \cite{NIPS2017_7062}.\begin{figure}
  \begin{subfigure}[t]{.48\textwidth}
    \centering
    \includegraphics[width=\linewidth]{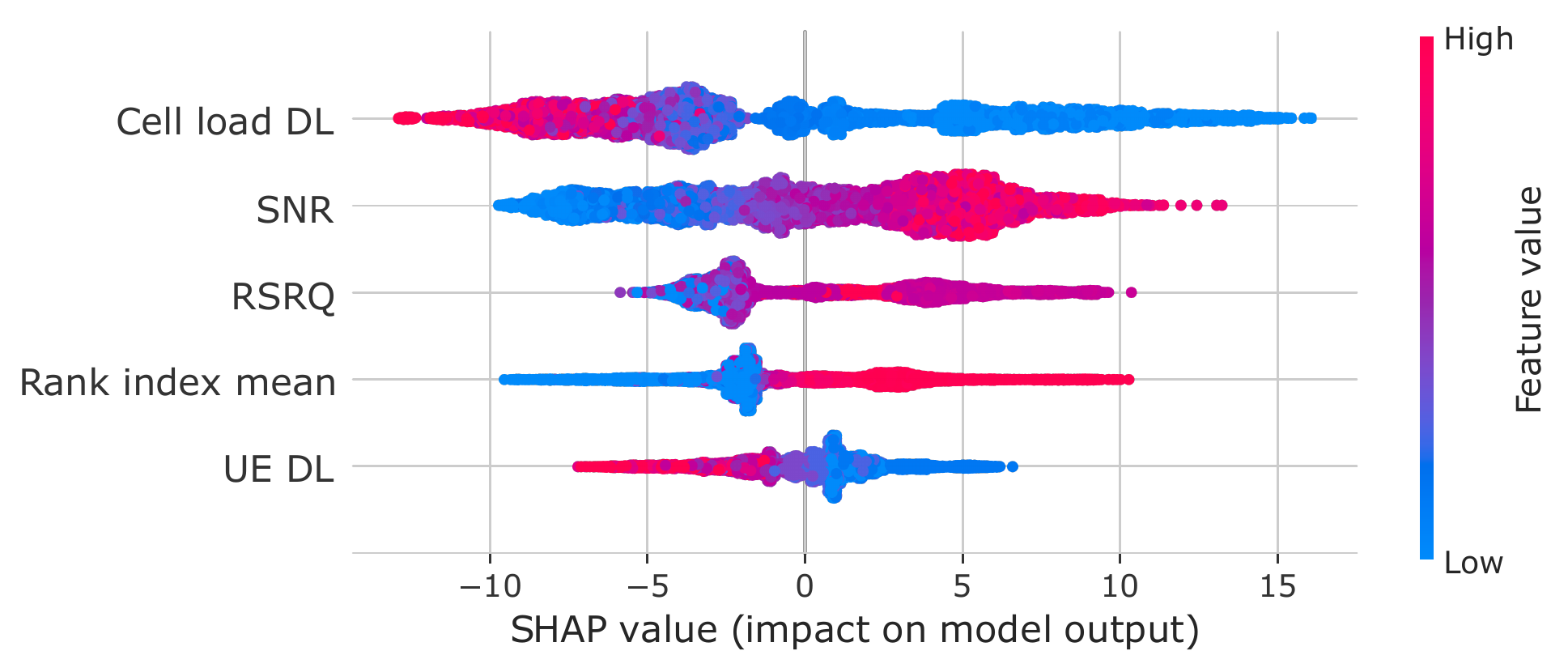}
    \caption{DL}
        \label{fig::shap_beea}
  \end{subfigure}
  \hfill
  \begin{subfigure}[t]{.48\textwidth}
    \centering
    \includegraphics[width=\linewidth]{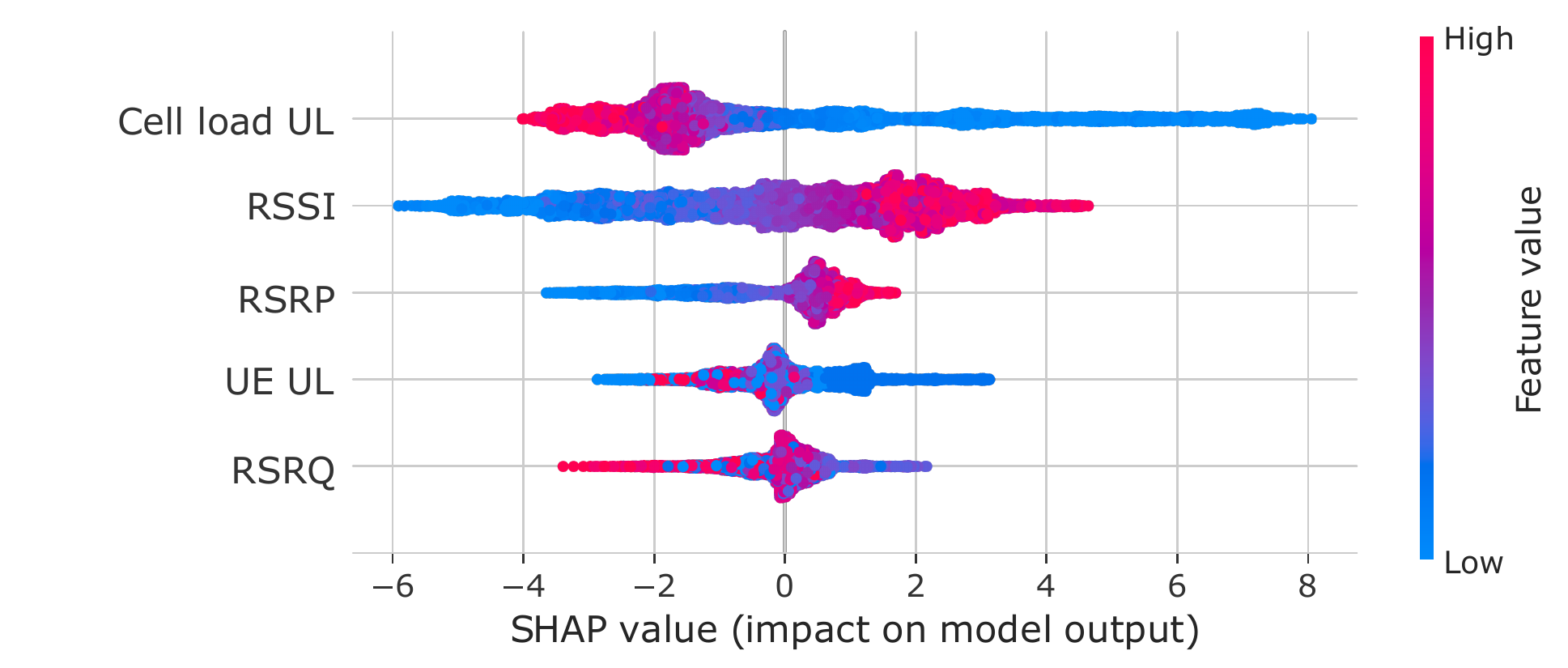}
    \caption{UL}
    \label{fig::shap_beeb}
  \end{subfigure}
    \caption{Feature importance in \gls{dll} and \gls{ul} of the five most important features in term of \gls{SHAP} values.}
      \label{fig::shap_bee}
\end{figure}
Fig. \ref{fig::shap_bee} presents the  \gls{SHAP} values of the five most important features, in  \gls{dll} and \gls{ul} respectively. The x-axis depicts the \gls{SHAP} value that describes the impact of the input feature on the prediction. The y-axis depicts the input feature names, and the color represents the numeric input feature value. 

The cell load feature has both in \gls{ul} and \gls{dll} a strong impact on the prediction. A higher cell load value leads to a lower value on the prediction and vice versa. On the other hand, the radio-based features (\gls{rsrp}, \gls{rsrq} and \gls{sinr}) seem to have the opposite tendency. A higher radio value leads to a higher value on the prediction and vice versa.

\subsection{Accumulated Local Effects}

We then continue using \gls{ALE} for discovering the global effects of input features.
\begin{figure}[htbp]
\centering
    \includegraphics[width=\columnwidth]{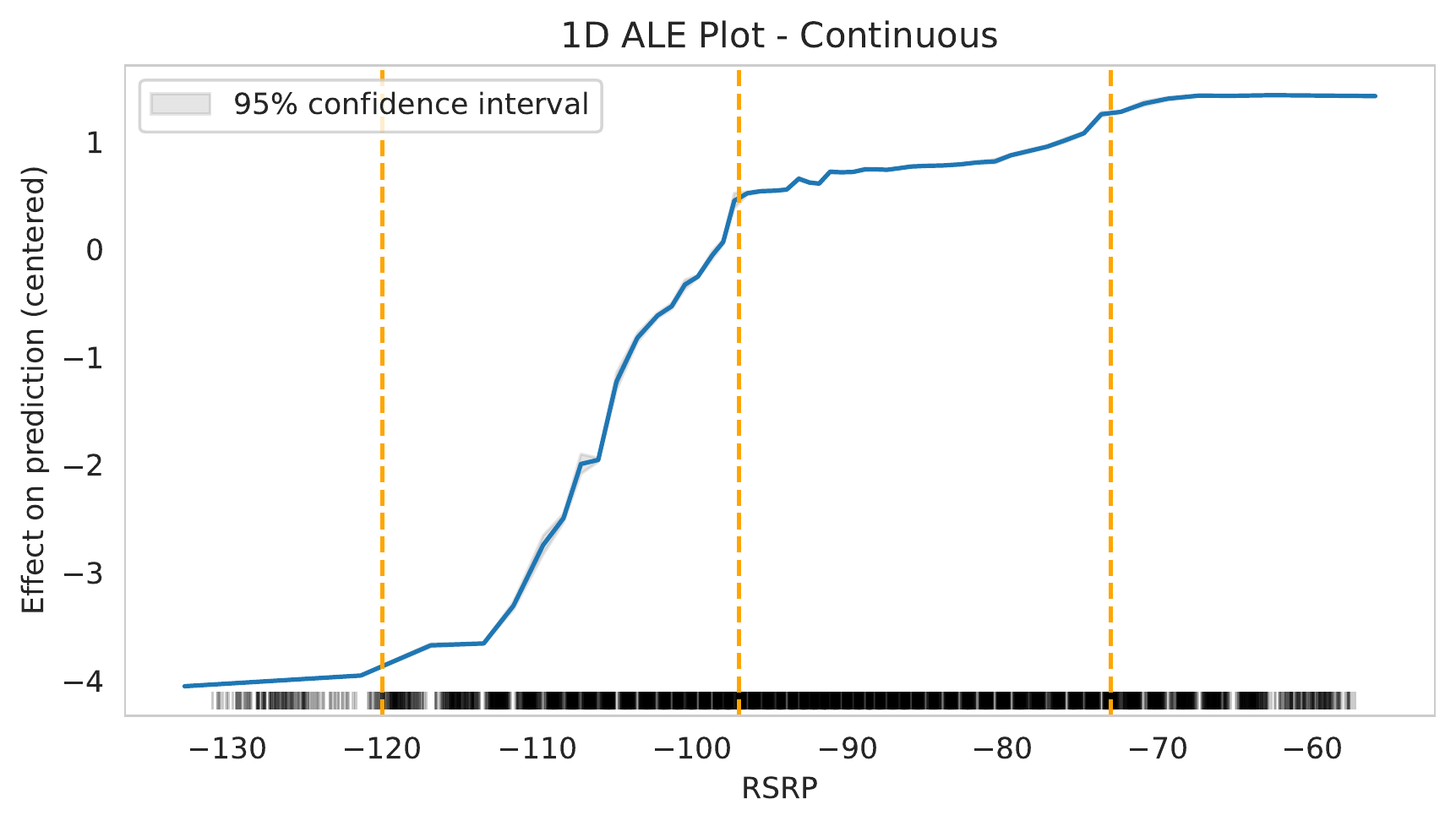}
\caption{ALE values for \gls{ul} \gls{rsrp}.}
    \label{fig::ale}
\end{figure}

Fig. \ref{fig::ale} depicts \gls{ALE} values for \gls{rsrp} in \gls{ul} throughput prediction. 
The x-axis depicts the \gls{rsrp} value, the y-axis the effect on the prediction, where a higher input feature value leads to a higher value of the prediction.

We note that in the ALE plot seems that there are four regions affecting the prediction differently as these are depicted in the figure. We added orange dotted lines at the perceived border of each region. 
The first one is with the lowest captured \gls{rsrp} values. The \gls{ue} needs to get at least some minimum \gls{rsrp} before it can transmit data. The second region shows some linear characteristics.  Higher \gls{rsrp} values seem to contribute to a higher throughput significantly in this range, which demonstrates that the \gls{rsrp} is contributing strongly to the prediction variable (\gls{ul} throughput). 
The third region shows some slight saturation in the linear trend, meaning that higher \gls{rsrp} values do not contribute considerably more to a higher throughput prediction. The last region looks saturated. This probably means that either higher \gls{rsrp} values do not provide any benefits for higher rates, or other factors like the cell load play a more important role on the prediction variable.  
We note that the \gls{ml} model has discovered these regions from the input features, without being explicitly programmed, which is also close to the operating principles of the cellular networks.

The interested reader should note that the absolute numbers of the regions depend to a large extent on the characteristics of the network deployment that include for example the  dynamic range of the receivers and the  network states captured. Similar type of learnings have been shown in a previous paper \cite{roman}, which was based on data from an operator in Asia. This increases the confidence that \gls{ml} is able to learn about wireless environment characteristics on diverse networks and deployments.

\section{Conclusion}\label{sec::conclusion}
Based on a dedicated measurement campaign, we have presented insights into building reliable \gls{ml} models for \gls{pqos}. Our results go beyond UE data by including network and vehicular information measurements, covering a large range of scenarios.
Our measurements reveal many challenges \gls{ml} models will face in real deployments. 

Our first contribution discussed methods for improving sampling for the radio environment. Its correlation structure allows improved sampling procedures that reduce energy consumption and signaling for sharing collected data. The vehicle speeds do not seem to impact the statistics and characteristics of the collected data strongly, further simplifying the sampling procedures. We have also tested the data stationarity assumption, a precondition for many \gls{ml} models and other theoretical approaches. Even though stationarity often holds true, especially for datasets captured over a longer duration, it should not be expected liberally. We have provided multiple examples where this assumption is violated, by focusing mostly on the aspect of concept drifts as vehicles move between radio environments. As concept drifts degrade \gls{ml} performance, a reactive \gls{ml}  method might be needed for retraining the \gls{ml} models as required, once drifts are detected.
Another option is having larger datasets that cover multiple scenarios, like in our testbed, reducing the number of concept drifts that an \gls{ml} model will face when deployed. A combination of the two methods might be the most feasible way forward. 

We were able to predict the maximum achievable throughput under high load scenarios, where multiple users compete for maximum throughput in the network, with an MAE of 2.46 and 1.08 Mbps for down- and uplink. In particular, our findings show that the effects of data processing, different validation datasets, and sets of input features can have a very strong overall effect on the \gls{ml} model performance. For that reason, simply comparing numbers from literature can be misleading. We have seen that \glsfirst{gb} models outperform neural networks while keeping a balance between complexity and explainability.
At the same time, our results clearly show that low-cost consumer-grade devices can be part of \gls{ml} processes. Their performance falls close to more expensive transceiver chains, further facilitating data collection procedures from end-user's terminals. 

Moreover, we emphasized the topics of interpretability and explainability, showing that the tested \gls{ml} models were able to capture the underlying principles without being explicitly programmed. It is interesting to note that the cell load was discovered as the most important feature for both communication directions, showcasing the importance of network features for such prediction tasks.

Our results indicate that more thorough testing of \gls{ml} models is needed as complexities coming from the radio environment, the end users, and the effects of the network can considerably affect prediction performance, which might not always be precisely captured by any collected dataset. That can result in high performance variations of \gls{ml} models, when deployed, that have never seen such effects in their training sets.

One can also draw several other conclusions and lessons from this data and its analysis, for the wireless research community and \gls{ml} engineers alike. First, reporting only the \gls{ml} performance of models on a specific dataset might provide overly optimistic results. Second, the data collection procedures, the handling, and processing of the data, as well as the way of reporting results, are all equally important in the long chain of \gls{ml} workflows. Third, results coming from simplified analytical assumptions and simulations should be used with caution. Our results indicate that real-world-captured intricacies might hinder the further performance of \gls{ml} models
than hitherto understood. 

Although some of these results could be regarded as intuitive, they have not been properly emphasized or taken into account in the literature, with only a few exceptions. 
Especially when it comes to applying \gls{ml} in wireless networks, there has been a tendency to rate the usefulness of \gls{ml} models based on their performance on some testing datasets. Our results clearly show that this can often be misleading, and more conservative estimates might need to be used. Moreover, the relevant acquisition costs of data are rarely discussed, and we believe that such aspects need to be part of future discussions and proposals. We hope that this publication serves as a starting point in that direction.

We believe that \gls{ml}-based predictions do not only have the potential to improve specific use cases but also serve as an important enabler for a more proactive network. We made more datasets \cite{hernangomez2022berlin, hernangomez2023towards} available to the research community that are based on similar collection principles as the ones we described.
In the future, we plan to integrate the lessons learned towards methods that can innately handle some of the dynamics we noticed in the radio environment, such as non-stationarities and concept drifts. Moreover, we would like to extend our work toward other \gls{kpi} metrics, such as latency and the number of dropped packets.  Finally, we would like to integrate issues of data governance since, in this study, we did not consider the acquisition cost of the different features in detail.

\ifCLASSOPTIONcaptionsoff
  \newpage
\fi



%
\bibliographystyle{IEEEtran}  
\bibliography{IEEEabrv,bibliography/main}  

%




\end{document}